\PassOptionsToPackage{table}{xcolor}
\documentclass[nonacm, acmlarge]{acmart}

\setcopyright{rightsretained}
\acmJournal{TOSEM}
\acmYear{2025} \acmVolume{1} \acmNumber{1} \acmArticle{1} \acmMonth{1} \acmPrice{}\acmDOI{10.1145/3767157}

\usepackage{amsmath,amsfonts}
\usepackage{algorithmic}
\usepackage{graphicx}
\usepackage{textcomp}
\usepackage[most]{tcolorbox}
\usepackage{layouts}
\usepackage{printlen}
\usepackage{pgfplots}
\usepackage{url}
\usepackage{array}
\usepackage{float}
\usepackage{pgfplots}
\usepackage{hyperref}
\usepackage{longtable}
\usepackage{listings}
\usepackage{supertabular}
\usepackage{subcaption}
\usepackage{booktabs}
\usepackage{threeparttable}
\usepackage{xcolor}
\definecolor{LightGray}{gray}{0.9}
\definecolor{LightBlue}{RGB}{220,230,240}
\definecolor{LightGreen}{rgb}{0.88, 1, 0.88}

\usepackage[font=small,labelfont=bf]{caption}
\newtcolorbox{myquote}[1][]{%
    colback=black!5,
    colframe=black!5,
    notitle,
    sharp corners,
    borderline west={2pt}{0pt}{black!80!black},
    enhanced,
    breakable,
}

\graphicspath{{figures/}}

\def\BibTeX{{\rm B\kern-.05em{\sc i\kern-.025em b}\kern-.08em
    T\kern-.1667em\lower.7ex\hbox{E}\kern-.125emX}}

\usepackage{todonotes}

\usepackage{csquotes}

\pgfplotsset{compat=1.18}

\begin{document}

\title{How do Machine Learning Models Change?\\
}
\titlenote{Author’s version. Version of Record: 
\href{https://doi.org/10.1145/3767157}{10.1145/3767157}.}

\author{Joel Castaño}
\email{joel.castano@upc.edu}
\affiliation{%
  \institution{Universitat Politècnica de Catalunya}
  \city{Barcelona}
  \country{Spain}
}

\author{Rafael Cabañas}
\email{rcabanas@ual.es}
\affiliation{%
  \institution{Department of Mathematics  and CDTIME,
University of Almería}
  \city{Almería}
  \country{Spain}
}

\author{Antonio Salmerón}
\email{antonio.salmeron@ual.es}
\affiliation{%
  \institution{Department of Mathematics and CDTIME,
University of Almería}  \city{Almería}
  \country{Spain}
}

\author{David Lo}
\email{davidlo@smu.edu.sg}
\affiliation{%
  \institution{Singapore Management University}
  \city{Singapore}
  \country{Singapore}
}

\author{Silverio Martínez-Fernández}
\email{silverio.martinez@upc.edu}
\authornote{Silverio Martínez-Fernández is the corresponding author.}
\affiliation{%
  \institution{Universitat Politècnica de Catalunya}
  \city{Barcelona}
  \country{Spain}
}

\begin{abstract}
The proliferation of Machine Learning (ML) models and their open-source implementations has transformed Artificial Intelligence research and applications. Platforms like Hugging Face (HF) enable this evolving ecosystem, yet a large-scale longitudinal study of how these models change is lacking. This study addresses this gap by analyzing over 680,000 commits from 100,000 models and 2,251 releases from 202 of these models on HF using repository mining and longitudinal methods. We apply an extended ML change taxonomy to classify commits and use Bayesian networks to model temporal patterns in commit and release activities. Our findings show that commit activities align with established data science methodologies, such as the Cross-Industry Standard Process for Data Mining (CRISP-DM), emphasizing iterative refinement. Release patterns tend to consolidate significant updates, particularly in model outputs, sharing, and documentation, distinguishing them from granular commits. Furthermore, projects with higher popularity exhibit distinct evolutionary paths, often starting from a more mature baseline with fewer foundational commits in their public history. In contrast, those with intensive collaboration show unique documentation and technical evolution patterns. These insights enhance the understanding of model changes on community platforms and provide valuable guidance for best practices in model maintenance.
\end{abstract}



\maketitle
\pagestyle{plain}     
\thispagestyle{plain}
\setlength{\headheight}{19.08403pt}

\section{INTRODUCTION}
The rapid advancement of Machine Learning (ML) has led to an extensive proliferation of open-source ML models (hereafter referred to as "models"), fundamentally transforming the landscape of Artificial Intelligence (AI) research and applications. Platforms such as Hugging Face (HF) \cite{HuggingFaceInc.2023} have become pivotal in this transformation by enabling the development, sharing, and deployment of models. These platforms foster a collaborative and dynamic ecosystem where researchers and practitioners continuously contribute to and refine a vast repository of models.

While existing research has delved into various facets of model maintenance, ranging from technical debt \cite{Bogner2021, Tang2021}, library usage \cite{Dilhara2021}, to architectural frameworks \cite{Leest2023}, there remains a notable gap in categorizing and analyzing the changes made to models over time. Specifically, no prior study has applied a multifaceted taxonomy of changes to ML repositories to systematically understand how these models are maintained and improved in practice.  This analysis is essential because models encompass unique elements such as data preprocessing, model parameters, training pipelines, and deployment configurations, which differ significantly from traditional software systems. Consequently, insights derived from general software evolution studies may not adequately capture the distinct and complex nature of model changes, highlighting the need for specialized frameworks and methodologies tailored to the ML ecosystem.

Understanding how models change over time is critical for several reasons:
\begin{itemize}
    \item \textbf{Maintenance and Operational Sustainability:} Continuous maintenance ensures that models remain functional and relevant as their dependencies and deployment environments evolve. This includes updating libraries, addressing compatibility issues, and ensuring models perform reliably in different operational contexts.
    \item \textbf{Improvement and Optimization:} By analyzing the nature of changes, developers can identify patterns that lead to more effective model improvements and optimizations.
    \item \textbf{Collaboration and Development Standards:} Insights into commit patterns and changes can inform better collaboration practices and help in establishing standardized workflows, coding conventions, and documentation practices. This fosters a cohesive development environment, enabling teams to work more effectively and maintain high-quality model development processes.
\end{itemize}

Building upon the ML change taxonomy introduced by Bhatia et al. \cite{bhatia2023towards}, which extends traditional software change classifications to capture the unique aspects of ML system development, this study aims to provide a large-scale analysis of how models change within the open-source ecosystem, focusing on HF. Bhatia et al.'s taxonomy introduces ML-specific change categories such as \textit{Model Structure}, \textit{Parameter Tuning}, and \textit{Training Infrastructure}, among others. By applying this taxonomy, we classify commits across 100,000 models and employ Bayesian networks (BNs) to uncover sequential patterns between commit and release activities over time. Our research addresses three main aspects to deliver a nuanced understanding of model changes:
\begin{itemize}
 \item \textbf{Categorization of Commit Changes:} We apply the ML change taxonomy to classify over 680,000 commits on HF, providing a detailed breakdown of change types and their distribution across models.
 \item \textbf{Analysis of Commit Sequences:} Utilizing BNs, we examine the sequence and dependencies of commit types to identify temporal patterns and common progression paths in model changes.
 \item \textbf{Release Analysis:} We investigate the distribution and evolution of release types, analyzing how model attributes and metadata change across successive releases, thereby shedding light on versioning and release practices within the HF ecosystem.
\end{itemize}

The structure of this paper is organized as follows. Section \ref{sec:background} presents the background and necessary concepts for understanding our study. Section \ref{sec:related_work} reviews related work, encompassing taxonomies of changes in software and ML systems and repository mining and longitudinal studies in software engineering and ML. Section \ref{sec:methodology} details the methodology, including the research goals, dataset construction, data preprocessing, commit classification process, and data analysis techniques employed to address the research questions. Section \ref{sec:results} showcases the results of our analysis, providing insights into file changes in ML repositories and addressing each of the research questions comprehensively. Section \ref{sec:discussions} discusses the implications of our findings, highlighting their significance for researchers and practitioners, and explores best practices for model development and maintenance. Section \ref{sec:threats} discusses threats to validity. Finally, Section \ref{sec:conclusions} concludes the paper by summarizing the key contributions and suggesting avenues for future research.

\textbf{Data availability statement}: Our replication package, including the datasets, code, and detailed documentation, is available on Zenodo \cite{zenodo}. The package is organized into folders for data collection, preprocessing, and analysis, each containing Jupyter notebooks and necessary scripts. Users can refer to the included README.md file for step-by-step instructions on setting up the environment, running the data extraction and preprocessing workflows, and executing the analysis notebooks to reproduce the results presented in the paper.

\section{BACKGROUND}\label{sec:background}

This section lays out the foundational concepts and technical context essential for understanding our study. We start by examining version control mechanisms in both traditional software and ML repositories, emphasizing how version control practices are adapted to meet the unique requirements of models on platforms like HF. Next, we explore BNs and Dynamic Bayesian networks (DBNs), clarifying their roles and advantages in modeling temporal dependencies and probabilistic relationships within software engineering and ML development processes.

\subsection{Version Control in Traditional Software and ML Repositories}

In the development and maintenance of models on platforms like HF, understanding the mechanisms of version control is essential. A \textit{commit} represents a set of changes applied to a model repository at a particular point in time, capturing updates to model code, configurations, or documentation \cite{loeliger2012git}. Conversely, a \textit{release} denotes a stable version of the model that is packaged and made available for deployment or public use \cite{sommerville2015software}. Releases typically encapsulate a collection of commits that introduce significant enhancements, fixes, or new features, and are often accompanied by release notes that summarize the changes \cite{github2024}. This structured approach to versioning facilitates the tracking of model evolution, ensures reproducibility, and supports collaborative development within the ML community.

Platforms such as GitHub adopt the concept of releases by associating them with Git tags, which mark specific points in the history of a repository. GitHub releases are snapshots of the repository at a specific tag, packaged and distributed to users along with release notes that describe the changes made \cite{github2024}. Tags allow developers to record significant milestones or versions in their software's evolution, ensuring that specific versions of the codebase can be retrieved and used in production.

In the context of ML repositories, platforms such as HF also supports a form of release management. HF repositories utilize Git branches and tags to store and mark different versions of models, datasets, or spaces. For example, a developer might tag a version of a model with \texttt{v1.0} to signify the model's release for public use, ensuring users can always reference and use that specific version. These tags, akin to traditional software releases, provide a mechanism to mark stable or significant versions of models that can be shared or deployed. The HF Hub API offers tools such as \texttt{list\_repo\_refs()} to manage and list all branches and tags within a repository \cite{huggingface2024}, ensuring users can access specific versions of models. This mirrors traditional software release mechanisms but is tailored to the unique needs of the ML community.

\begin{figure}[h!]
    \centering
    \begin{subfigure}[b]{0.48\textwidth}
        \centering
        \includegraphics[width=\textwidth]{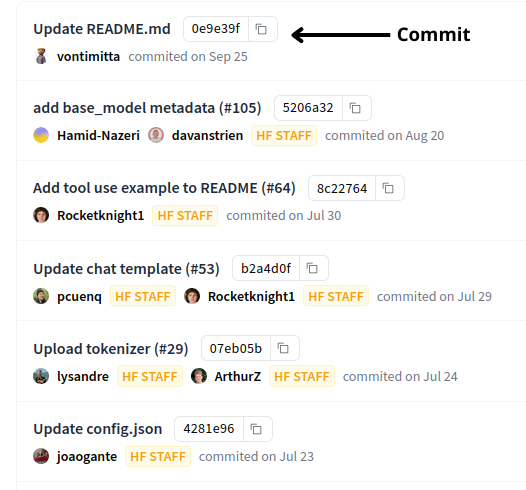}
        \label{fig:commit_example}
    \end{subfigure}
    \begin{subfigure}[b]{0.2\textwidth}
        \centering
        \includegraphics[width=\textwidth]{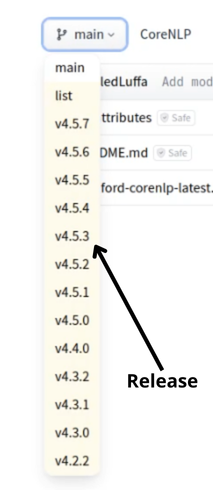}
        \label{fig:releases_example}
    \end{subfigure}
    \caption{Examples of commits and releases in HF model repositories.}
    \label{fig:commits_and_releases}
\end{figure}

To illustrate how commits and releases are managed in HF, consider Figure \ref{fig:commits_and_releases}:

\begin{itemize}
    \item The left image illustrates the commit history of a model repository (in this case, \texttt{Llama-3.1\allowbreak -8B-Instruct}). The commits include a variety of changes such as modifications to the external documentation (\texttt{README.md}), updates to the tokenizer, and alterations to configuration files like \texttt{config.json}. This diversity in commits highlights how granular changes are meticulously tracked over time, facilitating transparency and accountability in model development.

    \item The right image depicts the branches and tags within a model repository (in this case, \texttt{standford/CoreNLP}). The presence of multiple tags (e.g., \texttt{4.2.2}, \texttt{4.3.0}) indicates different stable versions of the model. These tags enable users to reference specific versions, ensuring that deployments are consistent and reproducible. The branching structure further supports parallel development and maintenance of different model versions.
\end{itemize}

This overview provides a foundation for understanding how version control mechanisms are applied within the ML ecosystem.

\subsection{Bayesian Networks and Dynamic Bayesian Networks in Software Engineering}

BNs are probabilistic graphical models that represent a set of variables and their conditional dependencies via a directed acyclic graph (DAG) \cite{pearl2014probabilistic,koller2009probabilistic}. In a BN, nodes represent random variables, and edges represent probabilistic dependencies between these variables. BNs provide a compact representation of joint probability distributions and are widely used for reasoning under uncertainty in various fields, including software engineering \cite{jensen2007bayesian}.

DBNs extend BNs to model temporal processes by representing sequences of variables over time~\cite{murphy2002dynamic}. In a DBN, the temporal evolution of a set of variables is modeled by replicating the network structure over multiple time steps, with connections between variables at different time steps capturing temporal dependencies. This allows DBNs to model complex stochastic processes where the state of the system evolves over time, making them suitable for time series analysis and sequence modeling.

In software engineering, BNs and DBNs have been utilized for various purposes, such as defect prediction~\cite{okutan2014software}, process modeling~\cite{tosun2017systematic}, and performance analysis~\cite{del2023bayesian}. BNs can capture the probabilistic relationships among software metrics, defects, and development practices, providing insights into software quality and project risks.

\subsubsection{Why DBNs for Modeling Commit Sequences}

In the context of our study, we aim to analyze the temporal patterns and dependencies in commit sequences to understand how different types of changes evolve over time in ML model repositories. DBNs are well-suited for this task due to several reasons:

\begin{itemize} \item \textbf{Temporal Modeling}: DBNs explicitly model temporal dependencies between variables at different time steps, allowing us to capture how the occurrence of a certain commit type at one time influences the likelihood of other commit types in subsequent times~\cite{murphy2002dynamic}. \item \textbf{Probabilistic Inference}: DBNs enable probabilistic reasoning about sequences, accommodating uncertainty and variability in commit behaviors~\cite{koller2009probabilistic}. \item \textbf{Handling Missing Data}: DBNs can handle incomplete data gracefully, which is common in real-world datasets where not all variables are observed at every time step~\cite{ghahramani1997learning}. \item \textbf{Scalability}: Efficient algorithms exist for learning the structure and parameters of DBNs from data, even for large datasets~\cite{friedman2013learning}. \end{itemize}

\subsubsection{Statistical Properties Suitable for This Study}

DBNs offer several statistical properties that make them appropriate for modeling commit sequences in our study:

\begin{itemize} \item \textbf{Markov Assumption}: DBNs typically assume that the state at time $t$ depends only on a limited history (e.g., the previous time step), which simplifies modeling and computation~\cite{murphy2002dynamic}. \item \textbf{Parameter Learning}: The parameters of a DBN can be learned from data using maximum likelihood estimation or Bayesian methods, allowing us to infer the strengths of dependencies between commit types~\cite{ghahramani1997learning}. \item \textbf{Structure Learning}: Algorithms for structure learning can identify the network topology that best explains the observed data, revealing the causal relationships between variables~\cite{scanagatta2019survey}. \item \textbf{Inference Efficiency}: Exact and approximate inference algorithms enable us to compute probabilities of interest efficiently, even in complex networks~\cite{koller2009probabilistic}. \end{itemize}

Given these properties, DBNs provide a powerful framework for uncovering patterns in the evolution of commits and releases over time. By modeling the dependencies between different commit types and project characteristics across time steps, we can gain insights into the dynamics of model development and maintenance in ML repositories.

\section{RELATED WORK}\label{sec:related_work}

In this section, we review existing literature that intersects with our research focus, encompassing taxonomies of changes in software and ML systems as well as empirical studies on ML repositories. We examine prior work on automated classification of code changes, repository mining studies specific to platforms like HF, and longitudinal analyses in software development practices.

\subsection{Taxonomies of Changes in Software and ML Systems}

Taxonomies of changes in software systems have evolved since Swanson's seminal work \cite{swanson1976dimensions}, which identified corrective, adaptive, and perfective changes during software maintenance. \citet{hindle2008large} extended these categories, providing a foundation for subsequent research. However, this taxonomy needed updates to accommodate the collaborative nature of modern software development and the specific requirements of ML systems.

\citet{bhatia2023towards} introduced a change taxonomy tailored for ML pipelines, expanding upon Hindle et al.'s framework by incorporating ML-specific change categories. Their taxonomy includes both traditional software engineering categories and ML-specific ones, introducing nine new subcategories such as \textit{Pre-processing}, \textit{Parameter Tuning}, \textit{Model Structure}, \textit{Training Infrastructure} or \textit{Pipeline Performance}. This comprehensive framework systematically captures the unique types of changes that occur in ML repositories, facilitating a nuanced analysis of model maintenance and evolution.

Recent studies have focused on automating the classification of code changes based on these taxonomies. For instance, \citet{hindle2009automatic} employed ML techniques to classify maintenance changes, which was subsequently improved by \citet{yan2016automatically} using Discriminative Probability Latent Semantic Analysis. Further advancements were made by \citet{ghadhab2021augmenting} who utilized BERT (Bidirectional Encoder Representations from Transformers) for enhanced classification. Additionally, \citet{li2016watch} developed a classification model to identify influential software changes early, achieving 86.8\% precision, 74\% recall, and 80.4\% F-measure. \citet{li2024understanding} demonstrated that small-scale language models, when fine-tuned on high-quality datasets, can effectively classify and summarize code changes, providing a cost-effective alternative to larger models.

Other notable contributions include \citet{janke20247}'s work on identifying context-specific code change patterns, \citet{dilhara2023pyevolve}'s tool for automating frequent code changes in Python ML systems, and \citet{dilhara2022discovering}'s fine-grained study on code change patterns in diverse ML systems.

Our study leverages Bhatia et al.'s taxonomy as the foundation for classifying commits within ML repositories on platforms like HF. By utilizing this established taxonomy, we ensure that our classification framework accurately captures both general software maintenance activities and ML-specific changes, providing a robust basis for analyzing model maintenance and evolution in the open-source ecosystem.

\newcommand{\customsizefirst}{\fontsize{8}{10}\selectfont}

\begin{table*}[ht]
\centering
\customsizefirst
\caption{Comparison of Repository Mining and Longitudinal Studies on ML Repositories}
\label{tab}
\begin{tabular}{|p{0.8cm}|p{2.5cm}|p{3cm}|p{3cm}|p{1cm}|}
\hline
 & \textbf{Object of Study} & \textbf{Examination Focus} & \textbf{Methodology} & \textbf{Year} \\
\hline
\rowcolor{LightGray}
\cite{Kathikar2023} & Open-source ML repositories (including HF) & Security vulnerabilities in open-source models & Repository Mining & 2023 \\
\hline
\rowcolor{LightGray}
\cite{Jiang2023} & Models in ML repositories & Practices and challenges of model reuse, dependency management & Mixed Methods (Qualitative Survey and Repository Mining) & 2023 \\
\hline
\cite{AIT2024103079} & HF platform & Development of \textit{HFCommunity}, a data collection tool & Engineering Research & 2024 \\
\hline
\cite{vsinikinteractive} & Models on HF & Introduction of interactive monitoring tool & Engineering Research & 2023 \\
\hline
\cite{gao2024documenting} & Models in ML repositories & Documentation of ethical aspects by developers & Qualitative Survey & 2024 \\
\hline
\rowcolor{LightGray}
\cite{pepe2023fairness} & Models in ML repositories & Fairness, bias, and legal issues & Repository Mining & 2023 \\
\hline
\rowcolor{LightGray}
\cite{jiang2023exploring} & Models in ML repositories & Naming conventions and defects, research-to-practice pipeline & Repository Mining & 2023 \\
\hline
\cite{jones2024we} & Model repositories & Systematic literature review and validation of qualitative claims & Systematic Review & 2024 \\
\hline
\rowcolor{LightGray}
\cite{yang2024ecosystem} & Models on HF & Ecosystem analysis of large language models for code & Repository Mining & 2024 \\
\hline
\rowcolor{LightGray}
\cite{castano2023analyzing} & Models on HF & Model development trends and maintenance practices & Repository Mining & 2023 \\
\hline
\rowcolor{LightGray}
\cite{ajibode2024towards} & Models on HF & Semantic versioning practices, naming conventions & Repository Mining & 2024 \\
\hline
\rowcolor{LightBlue}
\cite{harter2011does} & Software development practices & Relationship between process improvement and defect severity & Longitudinal Field Study & 2011 \\
\hline
\rowcolor{LightBlue}
\cite{fucci2018longitudinal} & Test-driven development practices & Retainment and effects on quality and productivity & Longitudinal Cohort Study & 2018 \\
\hline
\rowcolor{LightBlue}
\cite{carruthers2024longitudinal} & Software samples & Temporal validity of software samples & Longitudinal Study & 2024 \\
\hline
\rowcolor{LightGreen}
\textbf{This Study} & Models in repositories (focus on HF) & \textbf{Longitudinal analysis of commit and release patterns} & Repository Mining and Longitudinal Study & 2025 \\
\hline
\end{tabular}
\vspace{2mm}
\begin{tablenotes}
\item \textbf{Note:} Rows shaded in light gray represent studies using the \textit{Repository Mining} methodology, in light blue studies using the \textit{Longitudinal Study} methodology, in light green \textbf{our study}, which combines both methodologies, and in white other empirical studies.

\end{tablenotes}

\end{table*}

\subsection{Empirical Studies on ML Repositories: Repository Mining, Longitudinal Analyses, and Beyond}

Empirical studies on ML repositories have significantly advanced our understanding of model development and maintenance practices. Platforms such as HF, GitHub, and others host a plethora of models, providing rich data for various empirical investigations. \citet{Kathikar2023} conducted a security analysis across various ML repositories, including those linked to GitHub and HF, uncovering the presence of high-severity vulnerabilities in open-source models. This study highlights the complexities of securing models in open-source ecosystems. In our own previous work \cite{castano2023exploring}, we focused on the environmental impact of models hosted on platforms like HF, particularly their carbon footprint, underscoring the need for sustainable development practices in the ML community.

Recent studies have introduced tools and frameworks to facilitate the analysis of ML projects across different repositories. For instance, \citet{AIT2024103079} developed \textit{HFCommunity}, a tool that collects and integrates data from HF, emphasizing its growing role as a hub for collaborative development. Similarly, \citet{vsinikinteractive} introduced an interactive tool for monitoring the progress of open-source models on HF, providing insights into model architectures and author activities. \citet{yang2024ecosystem} analyzed the ecosystem of large language models for code on HF, identifying popular models and datasets, and highlighting practices in model reuse and documentation.

Researchers have also investigated various aspects of model reuse and maintenance in ML repositories. \citet{Jiang2023} explored the practices and challenges of model reuse, offering insights into dependency management in the ML ecosystem. \citet{pepe2023fairness} examined crucial aspects of fairness, bias, and legal issues associated with pre-trained models. \citet{gao2024documenting} investigated how developers document ethical aspects of open-source models, emphasizing the critical role of model documentation. \citet{jiang2023exploring} analyzed the naming conventions and defects of models, shedding light on the research-to-practice pipeline. Additionally, \citet{jones2024we} conducted a systematic literature review and validation of qualitative claims regarding model repositories. Our previous study \cite{castano2023analyzing} provided a view of model development trends and maintenance practices by exploring the community engagement, evolution, and maintenance of over 380,000 models hosted on HF.

Longitudinal studies have been instrumental in understanding the evolution and impacts of software development practices, tools, or technologies within specific communities or platforms. For example, \citet{harter2011does} investigated the relationship between software process improvement and defect severity, while \citet{fucci2018longitudinal} studied the retainment of test-driven development. \citet{carruthers2024longitudinal} analyzed the temporal validity of software samples in empirical software engineering research. In the specific context of model release practices on HF, \citet{ajibode2024towards} conducted an empirical study focused on semantic versioning (SemVer) adoption, naming conventions, and documentation transparency for pre-trained language models (PTLMs). Analyzing over 52,000 models, primarily through manual analysis and artifact comparison, their work revealed significant inconsistencies in versioning practices and highlighted challenges with implicit versioning (i.e., model changes without corresponding version updates).

Our study complements and significantly extends the work of \citet{ajibode2024towards}. While their research provides valuable insights into how PTLM releases are formally versioned and documented, our work takes a broader scope by analyzing both commits and releases across a wider range of model types on HF. Furthermore, our focus differs; we concentrate on classifying the *types* of changes occurring within commits and releases using an ML-specific taxonomy \cite{bhatia2023towards} and employ probabilistic modeling (BNs and DBNs) to understand the evolutionary dynamics and dependencies of these change types over time. Methodologically, we utilize automated LLM-based classification and analyze internal model metadata changes (RQ3.4), offering a different lens on model evolution compared to the versioning-focused analysis of Ajibode et al. Both studies, however, point towards inconsistencies and the need for greater transparency in the HF ecosystem, with Ajibode et al.'s findings on implicit versioning resonating with our observations of frequent commit activity between tagged releases.

Our study builds upon these diverse research areas to provide a large-scale analysis of the relationships and longitudinal patterns in model changes across repositories, with a focus on the HF platform as a case study. To clarify the contributions of prior work and position our study within the existing literature, we present a comparative table (Table~\ref{tab}) summarizing key aspects of the specified studies.

\section{METHODOLOGY}\label{sec:methodology}

\begin{figure}[!htp]
    \centering
    \includegraphics[width=\textwidth]{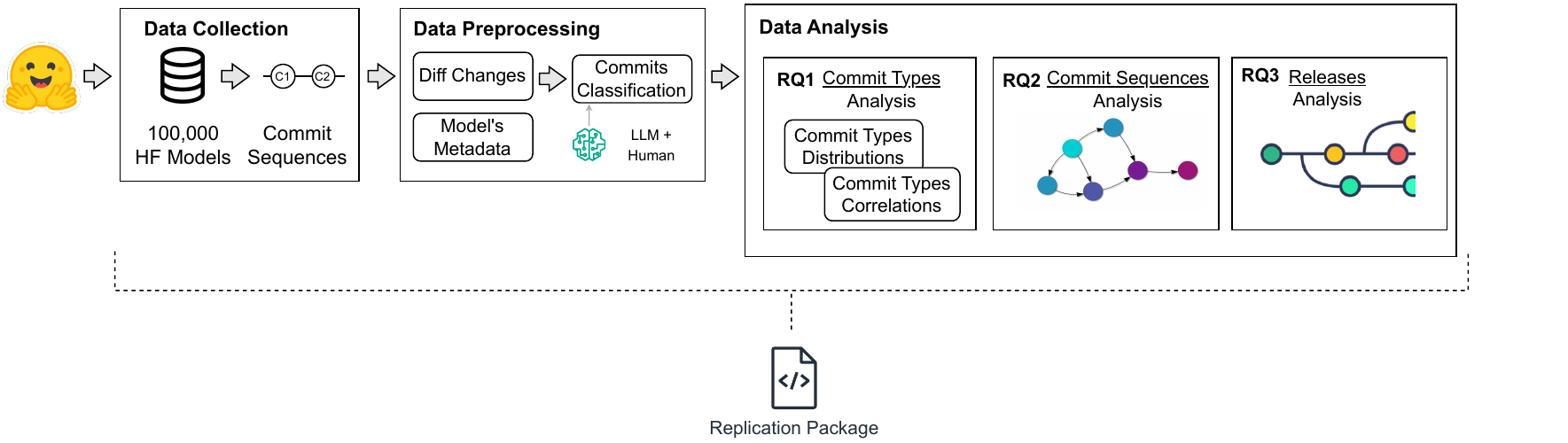}
    \caption{Data collection and analysis process}
    \label{study design evolution}
    \Description{Data collection and analysis process}
\end{figure}

In this section, we initially establish the objective of our study along with the research questions.  As illustrated in Fig.~\ref{study design evolution}, our study is structured into three primary phases. The first phase, \textit{Data Collection}, involves extracting data from the HF platform. This includes gathering commit histories, release information, and relevant metadata associated with the models. The second phase, \textit{Data Preprocessing}, encompasses the classification of commits and releases, as well as the cleaning and transformation of the collected data to prepare it for analysis. Finally, the third phase, \textit{Data Analysis}, utilizes the preprocessed data to address the research questions through various analytical techniques, including the application of BNs to uncover patterns and dependencies in the data. 

\subsection{Research Goal and Research Questions}

Following the Goal Question Metric (GQM) guidelines \cite{caldiera1994goal}, our research goal is structured as follows:

\textit{Analyze commits and releases of models for the purpose of classifying and understanding with respect to their changes from the viewpoint of ML researchers and practitioners in the context of open-source in the HF Hub.}

Understanding how models change over time is critical for ensuring their maintenance, sustainability, and continuous improvement. Despite the growing body of research on model development, there is a lack of comprehensive studies that systematically categorize and analyze the changes occurring within open-source ML repositories. This gap hinders the ability of practitioners to adopt best practices for model maintenance and optimization, and it limits researchers' understanding of the dynamics driving model changes. Therefore, our study aims to identify and characterize the patterns and factors influencing model changes on the HF platform.

To achieve this, we address the following research questions:

\begin{myquote} \textbf{RQ1}. \textit{How do models change based on the categorization of commit types?} \end{myquote}

\textbf{Motivation for RQ1}: Categorizing the types of changes made to models is essential for understanding the focus areas of development and maintenance efforts. By analyzing the distribution and evolution of commit types across the HF ecosystem, we can gain insights into common practices, identify potential areas for improvement, and understand how different project and commit characteristics are associated with these commit types. This knowledge can help practitioners prioritize resources and adopt best practices in model maintenance.

\begin{itemize} 
\item RQ1.1: What is the overall distribution and evolution of commit types across the HF ecosystem? 
\item RQ1.2: How are project and commit characteristics associated with commit types? 
\end{itemize}

\begin{myquote} \textbf{RQ2}. \textit{What are the patterns in the evolution of models based on commit sequences?} \end{myquote}

\textbf{Motivation for RQ2}: Beyond individual changes, the sequence and dependencies of commits provide valuable information about the development process and workflows. Analyzing commit sequences can reveal patterns in how different types of changes follow one another, dependencies between commit types over time, and how these dependencies are influenced by project characteristics. Understanding these patterns is crucial for optimizing development workflows, improving collaboration, and enhancing the efficiency of model evolution.

\begin{itemize} \item RQ2.1: What are the dependencies between different commit types over time? \item RQ2.2: How are these dependencies influenced by other commit and project characteristics? \end{itemize}

\begin{myquote} \textbf{RQ3}. \textit{How do models change based on the analysis of release versions?} \end{myquote}

\textbf{Motivation for RQ3}: Releases represent significant milestones in the development of models, encapsulating important changes and updates. Analyzing release types and the evolution of model attributes and metadata across successive releases can provide insights into versioning practices, the stability of models, and how significant updates are managed. This information is vital for practitioners to manage dependencies, ensure compatibility, and maintain the reliability of models in production environments.

\begin{itemize} \item RQ3.1: What is the overall distribution and evolution of release types across the HF ecosystem? \item RQ3.2: How are project and release characteristics associated with release types? \item RQ3.3: What are the patterns in the evolution of models based on release sequences? \item RQ3.4: How do model attributes and metadata change across successive releases? \end{itemize}

By addressing these research questions, our study aims to provide a large-scale understanding of model changes, enhancing model maintenance practices, optimizing development workflows, and fostering effective collaboration among ML researchers and practitioners in open-source communities.

\paragraph{\textbf{Methodological Approach}}

Bearing our research goal and research questions in mind, we adopted an approach that integrates both repository mining and longitudinal study methods, following guidelines recommended by the \textsl{ACM/SIGSOFT Empirical Standards}.\footnote{\url{https://github.com/acmsigsoft/EmpiricalStandards}.} These guidelines ensure that our analysis adheres to established best practices, enhancing the validity and reliability of our findings. Specifically, we maintain the identifiability of commits and releases over time, utilize appropriate statistical methods such as Bayesian networks to model dependencies, and address potential threats to validity through rigorous data preprocessing and validation procedures.

\textit{Repository Mining} was employed to quantitatively analyze a vast dataset extracted from the HF platform. This involved using automated tools to gather commit histories, release information, and relevant metadata from over 50,000 models. Repository mining is appropriate for our study as it allows us to handle and analyze large-scale data systematically, providing a broad overview of commit types and release patterns necessary to address RQ1.

\textit{Longitudinal Study} was utilized particularly for RQ2 and RQ3, to examine the temporal sequences and dependencies in commit and release activities over time. This method enables us to track the evolution of models, understanding how different types of commits and releases influence subsequent changes. By maintaining the identifiability of commits and releases across two different waves---namely, two consecutive commits at times \( t_0 \) and \( t_1 \), further explained at \ref{rq2_analysis}---the longitudinal study facilitates the analysis of how model development practices evolve. This approach aligns with our research questions, which focus on identifying patterns and changes over time.

Together, these methodologies provide a comprehensive framework for analyzing the dynamic nature of model changes on the HF platform. Repository mining offers the quantitative foundation for classifying and understanding commit types and release patterns, while the longitudinal study delves into the temporal dynamics and dependencies that drive the evolution of these models.

Next, we detail the method for acquiring the dataset necessary for the examination of these research questions.

\subsection{Dataset Construction}

To address our RQs, we build upon the dataset from our previous study \cite{castano2023analyzing},  which details the initial data collection process, including the extraction of commit histories, and relevant metadata from the HF platform. Here, we provide a summary overview of how this dataset was created and elaborate on the additional steps taken to extend this dataset for the current study.

\begin{figure}[!hbp]
    \centering
    \includegraphics[width=\textwidth]{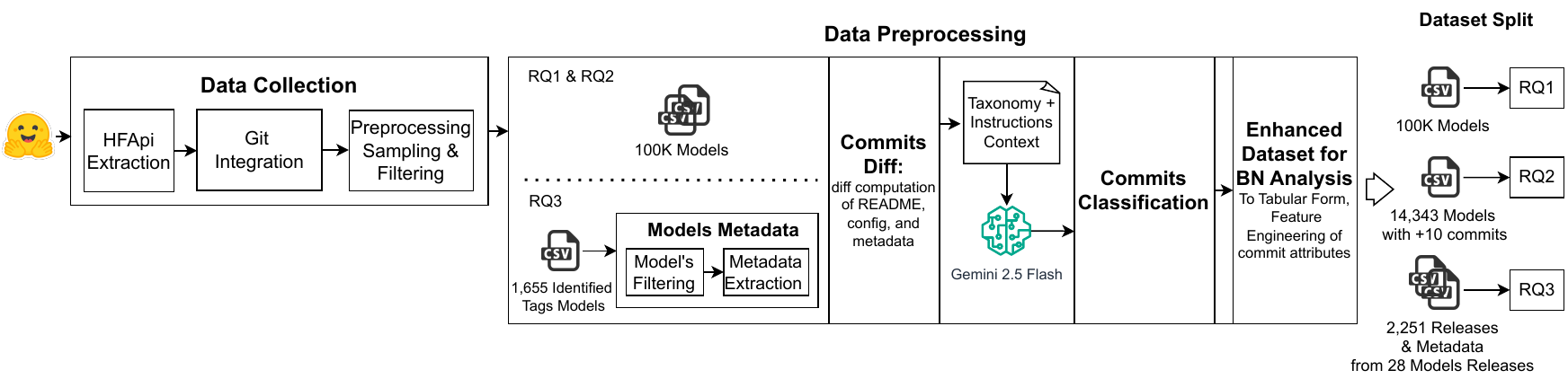}
    \caption{Dataset Construction Process}
    
    \label{dataset construction}
\end{figure}

The dataset construction process is illustrated in Fig. \ref{dataset construction}, which includes a comprehensive diagram outlining each step. As illustrated in the figure, the dataset construction methodology is divided into three main stages: \textit{Data Collection}, \textit{Data Preprocessing}, and \textit{Data Splitting}. In the \textit{Data Collection} stage, we gather relevant data from the HF platform, including commit histories, release information, and associated metadata for each model. The \textit{Data Preprocessing} stage involves, among others, classifying commits and releases according to our extended taxonomy, as well as cleaning and transforming the data to ensure consistency and reliability for subsequent analysis. Finally, the \textit{Data Splitting} stage entails dividing the preprocessed data into distinct subsets tailored to address each specific research question (RQ1, RQ2, and RQ3), thereby facilitating targeted and efficient analysis.

\subsubsection{\textbf{Data Collection}}

To comprehensively address our research questions, we constructed two distinct datasets from the HF platform: the \textbf{Commit Dataset} and the \textbf{Tag Release Dataset}. Each dataset serves a specific purpose in our analysis, enabling a nuanced examination of model changes and release patterns.

\textbf{Commit Dataset:} The foundation of this dataset originates from the data collection pipeline detailed in our prior work \cite{castano2023analyzing}. That work describes a process utilizing the HF Hub API and the \texttt{HfApi} class \cite{huggingfaceHfApiClient} to gather a wide array of information about models, including common attributes such as dataset sizes, training hardware, evaluation metrics, model file sizes, download and like counts, descriptive tags, model card texts, and basic commit history information available via the API. For the present study, data extraction is performed up to May 2025 to capture the most recent trends. A key enhancement in this study is the method to obtain a detailed list of files modified in each commit. While our previous work notes the utility of datasets like \textit{HFCommunity} \cite{Ait2023} (which itself used PyDriller for such details), for this research, to ensure recency and control over the extraction process, we implement a direct Git processing approach. For each model in our sample, its repository is temporarily cloned (using a Git bare clone with no checkout and blob filtering for efficiency). We then programmatically execute `git show --name-only` for every commit SHA obtained from the HF API's commit history for that model. This process, detailed in our replication package \cite{zenodo}, allows us to retrieve the list of changed files for each commit, data not consistently or fully available directly via the HF API for all historical commits.

From the total pool of 1,657,941 models available on HF (as of our data extraction in May 2025), we randomly sample 100,000 models for RQ1. This sample size strikes a balance between achieving high statistical power, providing a margin of error of approximately 0.3\% at a 95\% confidence level for estimating proportions near 50\%, and ensuring practical feasibility regarding computational resources, time constraints, and API usage limits \cite{cochran1977sampling}. This margin of error was calculated using the standard formula for a sample proportion, assuming a 95\% confidence level, a conservative population proportion of 0.5 (to maximize variance), our sample size of $n=100,000$ models, and applying a finite population correction. This large sample allows us to capture the diversity within the HF ecosystem, although it reflects the platform's known domain distribution, which is predominantly composed of NLP models \cite{jiang2024peatmoss}. To ensure the representativeness of typical development practices and mitigate the skewing effect of outlier repositories with an exceptionally high number of commits (e.g., automated or bulk updates), we applied a filter to exclude models with more than 500 commits. After this filtering, the total number of commits in our final dataset for analysis is 680,966. The potential implications of this characteristic on the generalizability of our aggregate findings are further discussed in Section \ref{sec:threats} (Threats to Validity).

For RQ2, which focuses on the sequence of commits, we further filter this 100,000-model sample to include only those models with at least 10 commits. The rationale for this threshold is to focus the evolutionary analysis on models with a discernible history of changes, excluding newly created, inactive, or minimally changed repositories, thereby ensuring sufficient longitudinal data points for studying temporal patterns. This filtering identifies 14,343 such models. This subset of actively evolving models, comprising all models from our initial sample that met the commit threshold, is then used for the DBN analysis in RQ2. This dataset size (14,343 models) is substantial for robust sequence modeling with DBNs. Moreover, as these models are identified from our large random sample (in which 14.34\% met the $\geq$10 commits criterion), they allow for generalizing findings about evolutionary patterns to the broader sub-population of active models on HF. This sub-population is estimated to be approximately 237,820 models (14.34\% of the 1,657,941 total models identified at the time of our data extraction in May 2025). For estimating proportions of characteristics within this active sub-population, our 14,343 models provide a margin of error of approximately 0.79\% at a 95\% confidence level, calculated using the same conservative assumptions (e.g., employing a population proportion of $p=0.5$) and finite population correction methodology as detailed for the RQ1 sample. Using this full subset allows for a comprehensive analysis of sequence modeling within this active population while managing computational demands.

\textbf{Tag Release Dataset:} In addition to commit data, we create a specific dataset focusing on model releases marked by Git tags. Tags in HF repositories signify specific states or versions of a model. We initially identify 1,655 models with any detectable Git tags by programmatically checking repository references (via the hf\_api.list\_repo\_refs API function) across our collected models up to May 2025. We acknowledge this number is relatively small compared to the total number of models on HF; this may be due to models following versioning implicitly (e.g., in READMEs or filenames) without explicit Git tags \cite{ajibode2024towards}, or varying adoption rates of formal tag-based versioning. Our subsequent analysis therefore focuses on the subset of models employing explicit Git tagging.

To refine this set for meaningful longitudinal analysis of release practices, we apply several filtering steps (detailed in our replication package \cite{zenodo}):
\begin{itemize}
\item Tags are filtered using a regular expression to retain only those likely representing semantic or calendar versions, excluding arbitrary tags (e.g., global\_step200).
\item We require models to have at least five such distinct, plausible release tags. The distribution of tags per model (before this length filtering) was highly skewed, with a median of 1 tag and 75\% of models having 6 or fewer tags. The minimum of five releases is chosen to provide a sufficient basis for analyzing sequences and evolution across multiple development stages, aligning also with our prior lifecycle analysis \cite{castano2023analyzing}.
\item An upper limit of 50 releases per model is applied to exclude a few extremely prolific repositories that might disproportionately skew the analysis or represent automated/bulk tagging rather than typical curated releases. This filtering process results in a dataset of 204 models. Due to occasional authorization issues in extracting full commit histories for all tags of these 204 models, our final dataset for RQ3.1, RQ3.2, and RQ3.3 comprises 202 models, encompassing a total of 2,251 releases. Our findings for these RQs are thus representative of this specific subset of models on HF that utilize explicit, systematic Git tagging for release management.
\end{itemize}

For the specific analysis of internal metadata changes within model files (RQ3.4), from these 202 models, we further select a subset of 28 models. This subsetting is necessary due to feasibility constraints. The primary constraints are: 1) \textit{Storage}: Downloading all releases for 202 models could require terabytes. 2) \textit{Processing Time/Compute}: Parsing thousands of potentially very large model files is computationally intensive. 3) \textit{Tooling Feasibility}: Our metadata extraction scripts were most reliable for common formats. Therefore, the subset of 28 models was selected based on releases using specific, programmatically parseable file extensions (\texttt{.bin}, \texttt{.pth}, \texttt{.pt}, \texttt{.ckpt}, \texttt{.safetensors}, \texttt{.h5}) from which detailed metadata can be reliably extracted. We deliberately excluded \texttt{.gguf}, despite its prevalence (see Section~\ref{sec:results_file_changes}), because it is an inference-oriented, quantized format that typically omits optimizer/training state and uses flat, implementation-specific tensor naming. While basic tensor headers can be read with dedicated GGUF parsers, reproducing our analysis (keys/shapes/dtypes \emph{and} optimizer states, plus mean/std statistics) would require a separate parsing path and streaming/dequantizing large tensors, exceeding our compute/memory budget and reducing comparability with our Torch/HDF5/safetensors pipeline. Extending our extractor to robustly support \texttt{.gguf} is left as future work. The resulting metadata dataset supports our analysis of internal parameter and configuration evolution over time (RQ3.4).

\subsubsection{\textbf{Data Preprocessing}}

To prepare the datasets for analysis, we perform two main preprocessing tasks: computing commit diffs for the Commit Dataset and extracting model metadata for the Tag Release Dataset.

\textbf{Commit Diffs:} We compute the differences between commits for key files, specifically .json files (e.g., config.json). For each commit that modifies these files, we compare the changes with the previous commit affecting the same file using the \textit{difflib} library. This allows us to identify added, deleted, and updated keys in the .json files, providing the granularity needed to classify the commit changes based on \citet{bhatia2023towards}'s taxonomy. README diff information is not considered because README differences are mostly related to documentation. Including them significantly increases the token overhead and can worsen performance by introducing a lot of redundant context.

\textbf{Model Metadata:} We extract detailed metadata from the model files of the releases of the 28 selected models, focusing on attributes such as:

\begin{itemize}
    \item Keys: Names of the parameters in the model (e.g., `bert.embeddings.word\_embeddings.weight').
    \item Shapes: Shapes of the tensors (e.g., `[30522, 768]').
    \item Data Types: Data types of the tensors.
    \item Total Parameters: Total num. of parameters in the model.
    \item Optimizer States: Keys that are optimizer states.
    \item Mean Values: Mean values of the tensors.
    \item Nested Structures: Nested structures in the model.
\end{itemize}

This list of attributes is selected to capture the primary structural and numerical characteristics programmatically accessible from standard model save files (like PyTorch state dictionaries) using common libraries (e.g., PyTorch, safetensors, h5py). The reasoning is to identify changes potentially related to: 1) \textit{Architecture} (via keys, shapes, nested structures), 2) \textit{Size/Complexity} (via total parameters), 3) \textit{Precision} (via data types), 4) \textit{Training State} (via optimizer states), and 5) \textit{Weight Values} (via high-level mean values, as detailed comparison was beyond scope). While this covers key aspects commonly inspected, it is not exhaustive (e.g., it doesn't capture activation functions used in code or specific hyperparameters if not saved in the state dictionary). However, it represents metadata readily and reliably extractable from the saved state files and is considered sufficiently comprehensive for identifying major shifts in structure or numerical properties between releases for RQ3.4.

The extraction is performed using a custom Python script leveraging these libraries. The high-level process involves loading the model file (e.g., using `torch.load` loaded onto CPU, `safetensors.load\_file`, `h5py.File`), accessing the state dictionary or tensor structure, recursively iterating through it, and programmatically extracting the metadata for each tensor. Aggregated information like the total parameter count is also computed. The extracted metadata for each release file is stored for comparison. Further details and the specific script (\texttt{HFReleasesPreprocessing.ipynb}) are available in our replication package \cite{zenodo}.

To analyze the evolution of these internal model characteristics for RQ3.4, we then compute the differences between the extracted metadata of successive releases for each of the 28 selected models. This process involves first sorting the releases chronologically for each model. Then, for each release (from the second one onwards), its metadata attributes are compared against those of its immediately preceding release. The comparison identifies:
\begin{itemize}
    \item Numerical differences in aggregate statistics like the total number of parameters and the count of layers/keys.
    \item Structural changes within the model's parameter dictionaries (or equivalent structures for H5 files). For attributes like parameter shapes, data types, mean values, standard deviations, and nested structures, we identify:
        \begin{itemize}
            \item Added keys: Parameters or layers present in the current release but not in the previous one.
            \item Removed keys: Parameters or layers present in the previous release but not in the current one.
            \item Changed key-value pairs: Parameters or layers present in both releases but with modified attributes (e.g., a tensor's shape, data type, or its mean/std value changing).
        \end{itemize}
\end{itemize}
This detailed diffing provides fine-grained insights into how model parameters, architectural elements (as reflected by keys and shapes), and numerical properties evolve from one tagged release to the next, forming the basis of our analysis for RQ3.4.

\newcommand{\customsize}{\fontsize{5.5}{6.5}\selectfont}

\begin{table*}[h!]
\centering
\caption{Commit Types Based on Bhatia et al.'s Taxonomy with Examples and Explanations}
\begingroup
\renewcommand{\arraystretch}{1.25} 
\customsize
\begin{tabular}{|>{\raggedright\arraybackslash}m{1cm}|>{\raggedright\arraybackslash}m{1.5cm}|>{\raggedright\arraybackslash}m{8cm}|>{\raggedright\arraybackslash}m{3cm}|}
\hline
\textbf{Type} & \textbf{Description} & \textbf{Example} & \textbf{Explanation} \\ \hline
Pre-processing & Internal data manipulation or transformation logic prior to model training. & \textbf{Model ID:} facebook/detr-resnet-50 \; \textbf{ID:} c437d7cfa4d43b5276efc53fef63b91398c6ab3d \newline \textbf{Title:} First commit \newline \textbf{Files modified:} config.json, preprocessor\_config.json \newline \textbf{Changes to config.json:} Added keys related to preprocessor configuration & The changes included adding keys related to preprocessor configuration in config.json, which deals with data manipulation before model training. \\ \hline
Parameter Tuning & Adjustments to hardcoded hyper-parameters within the ML pipeline. & \textbf{Model ID:} nuigurumi/basil\_mix \; \textbf{ID:} a171dc37afa8cf7fb5aff196f98fb1a69ea8722f \newline \textbf{Title:} Add scale\_factor to vae config. (\#10) \newline \textbf{Files modified:} vae/config.json \newline \textbf{Changes to vae/config.json:} Added key "scaling\_factor" & Adds a new hyper-parameter "scaling\_factor" to the VAE configuration, which is an adjustment to the model's hyper-parameters. \\ \hline
Model Structure & Structural changes to the model's code (e.g., model architecture modification). & \textbf{Model ID:} tiiuae/falcon-40b \;\textbf{ID:} 4a70170c215b36a3cce4b4253f6d0612bb7d4146 \newline \textbf{Title:} Move to in-library checkpoint (for real this time) (\#107) \newline \textbf{Files modified:} config.json \newline \textbf{Changes to config.json:} Added keys "num\_attention\_heads",  "num\_hidden\_layers"... & Includes changes to the model configuration such as "num\_attention\_heads", indicating a structural modification of the model. \\ \hline
Training Infrastructure & Changes affecting the model training logic. & \textbf{Model ID:} nitrosocke/mo-di-diffusion \; \textbf{ID:} 0f297645c9e8ec991ae1ba8eb7e9c4f1d7587619 \newline \textbf{Title:} Add clip\_sample=False to scheduler to make model compatible with DDIM. (\#20) \newline \textbf{Files modified:} scheduler\_config.json \newline \textbf{Changes to scheduler\_config.json:} Added key "clip\_sample" & Adds a key "clip\_sample" to the scheduler configuration, which affects the model training logic, specifically for compatibility with DDIM. \\ \hline
Pipeline Performance & Modifications enhancing ML run-time pipeline efficiency. & \textbf{Model ID:} THUDM/chatglm2-6b-int4 \; \textbf{ID:} 5579a9f4c07b1dde911efedfba78af372aacd93a \newline \textbf{Title:} Update quantized gemm kernel \newline \textbf{Files modified:} quantization.py \newline \textbf{Changes:} Update quantized gemm kernel & Updates the quantized GEMM kernel, which is a change aimed at enhancing the performance of the model's pipeline. \\ \hline
Sharing & Changes to project presentation or deployment to improve collaboration. & \textbf{Model ID:} mosaicml/mpt-7b \; \textbf{ID:} c5ccdb75f2dcb9f256204e52bdc30b8f98c8119b \newline \textbf{Title:} Upload folder using huggingface\_hub \newline \textbf{Files modified:} config.json \newline \textbf{Changes to config.json:} Added multiple keys for model configuration & Involves uploading files using the huggingface\_hub, which facilitates better sharing and collaboration within the community. \\ \hline
Validation Infrastructure & Modifications to components evaluating model performance. & \textbf{Model ID:} THUDM/chatglm-6b-int4 \; \textbf{ID:} 630d0efd8b49de29a5c263b5055926ec71980f50 \newline \textbf{Title:} Add assertion when loading CPU and CUDA kernel fails \newline \textbf{Files modified:} quantization.py \newline \textbf{Changes:} Added assertion when loading CPU and CUDA kernel fails & Adds assertions for CPU and CUDA kernel loading, enhancing the validation infrastructure of the model. \\ \hline
Internal Documentation & Explaining a pipeline's internal execution to developers, e.g., via log files. & \textbf{Model ID:} openbmb/cpm-bee-10b \; \textbf{ID:} 1b34eda1006c1b2aca6288ed33ac9a8f28ba511c \newline \textbf{Title:} add resource files \newline \textbf{Files modified:} config.json \newline \textbf{Changes to config.json:} Added keys related to model configuration & Includes additions to the model configuration, aiding internal documentation and understanding of code workings. \\ \hline
External Documentation & Changes to end-user documentation. & \textbf{Model ID:} databricks/dolly-v2-12b \; \textbf{ID:} 19308160448536e378e3db21a73a751579ee7fdd \newline \textbf{Title:} add citation \newline \textbf{Files modified:} README.md \newline \textbf{Changes:} Added citation information to README.md & Adds citation information to the README, which is a change to the external documentation for end-users. \\ \hline
Input Data & Logic for ingesting raw, external data sources. & \textbf{Model ID:} openai/clip-vit-large-patch14 \; \textbf{ID:} 2cea2ab5ae7bc10ab11bb8569513495d800f86f0 \newline \textbf{Title:} add tokenizer.json \newline \textbf{Files modified:} tokenizer\_config.json \newline \textbf{Changes to tokenizer\_config.json:} Modified key "special\_tokens\_map\_file" & Involves adding and modifying keys in the tokenizer configuration, affecting how input data is handled. \\ \hline
Output Data & Any artifact produced and saved by the ML pipeline (e.g., models, logs). & \textbf{Model ID:} anon8231489123/vicuna-13b-GPTQ-4bit-128g \; \textbf{ID:} d95d41022e5aaed996ec616dedf3eb7667c1e968 \newline \textbf{Title:} Safetensor added. Use this. \newline \textbf{Files modified:} vicuna-13b-4bit-128g.safetensors \newline \textbf{Changes:} Added safetensor file & Adds a new safetensor file, which changes how the output data is stored. \\ \hline
Project Metadata & Changes to repository configuration or metadata that identifies the project. & \textbf{Model ID:} THUDM/chatglm2-6b \; \textbf{ID:} d17f53d7183e917ce1dbd329ee30e0c98703b907 \newline \textbf{Title:} initial commit \newline \textbf{Files modified:} .gitattributes \newline \textbf{Changes:} Initial commit adding .gitattributes file & The initial commit includes adding the .gitattributes file, which is related to project metadata. \\ \hline
Add Dependency & Introduction of a new dependency. & \textbf{Model ID:} Writer/camel-5b-hf \; \textbf{ID:} 0a47f3a2545f165ea80d37822c2e1683ff25a518 \newline \textbf{Title:} Create requirements.txt (\#1) \newline \textbf{Files modified:} requirements.txt \newline \textbf{Changes:} Created requirements.txt & Introduces a new dependency by creating the requirements.txt file. \\ \hline
Remove Dependency & Removal of an existing dependency. & \textbf{Model ID:} philschmid/pyannote-speaker-diarization-endpoint \; \textbf{ID:} dd70eb1d1c526dbb30f50294041c5213320956ab \newline \textbf{Title:} Delete requirements.txt \newline \textbf{Files modified:} requirements.txt \newline \textbf{Changes:} Deleted requirements.txt & Removes an existing dependency by deleting the requirements.txt file. \\ \hline
Update Dependency & Updates to the metadata of an existing dependency. & \textbf{Model ID:} hakurei/waifu-diffusion \; \textbf{ID:} 87a6d830b9b23f7e5727f162782cf3f4a7a84be1 \newline \textbf{Title:} Fix deprecated float16/fp16 variant loading through new version API. (\#133) \newline \textbf{Files modified:} .gitattributes, safety\_checker/model.fp16.safetensors \newline \textbf{Changes:} Fixed deprecated float16/fp16 variant loading through new "version" API & The commit updates metadata for existing dependencies, ensuring compatibility with the new version API for float16/fp16 variant loading. \\ \hline
\end{tabular}
\endgroup

\label{table:commit_types}
\end{table*}

\paragraph{\textbf{Commit Classification}}

With all the diff information engineered and extracted, we proceed to classify each commit to determine its type of change according to Bhatia et al.'s taxonomy \cite{bhatia2023towards}, displayed in Table \ref{table:commit_types}.

Table \ref{table:commit_types} comprises four columns: \textit{Type}, \textit{Description}, \textit{Example}, and \textit{Explanation}. The \textit{Type} column lists the commit types from Bhatia et al.'s taxonomy, which are the labels assigned to commits by the LLM based on the input provided. The \textit{Description} column provides a brief explanation of each commit type. The \textit{Example} column presents an actual commit from our dataset, including the model ID, commit ID, commit title, files modified, and relevant changes. This example represents the input provided to the LLM for classification. The \textit{Explanation} column offers a rationale for why, given the input, it is reasonable that the LLM classified the commit into the specified \textit{Type}.


To classify the ML-specific commit types from Bhatia et al.'s taxonomy, we use the Gemini 2.5 Flash LLM \cite{deepmindGeminiFlash}. This model is chosen because, at the time of classification (May 2025), it provides an optimal balance of effectiveness, cost, and speed, achieving performance comparable to state of the art LLMs (e.g., LMSYS Chatbot Arena~\cite{huggingfaceLMSysChatbot}) at approximately 50 times reduced cost and faster inference \cite{gpt4Pricing,googleGeminiPricing}.

The large scale of commit classification also necessitates strategies for managing LLM API rate limits. Our approach includes selecting the Gemini 2.5 Flash model partly for its flexible token-based limits and implementing robust error handling in our classification scripts. Specifically, LLM API calls incorporated automatic retries with exponential backoff (e.g., pausing for 60 seconds or more) upon encountering rate limit errors (typically indicated by a 429 HTTP status code), ensuring that temporary interruptions did not halt the overall classification process.

For the classification process, we provide the LLM with context regarding the classification criteria (the descriptions of each commit type) and the commit data extracted previously. Specifically, the input to the LLM included details about each commit, such as:

\begin{verbatim}
Commit Number
Commit Title: {raw title}
Commit Message: {raw message}
Files modified during commit
{config_diff}
\end{verbatim}

We ensured that the output from the LLM was in JSON format to facilitate the retrieval and processing of the classifications for each commit.

To ensure the reliability and objectivity of our LLM-based classification, we implemented a rigorous two-phase validation process. Full details of the process, including the manual classifications and prompt iterations, are available in our replication package \cite{zenodo}.

\begin{enumerate}
    \item \textbf{Prompt Refinement Phase (Training):} We first developed and refined the LLM prompt using a curated training set of 143 commits, specifically selected to ensure adequate representation across all classification categories. To establish a reliable ground truth, a primary author coded all 143 commits. The consistency of this annotation was then verified by a second author who independently coded a subset of 40 commits (approx. 28\% of the set). This practice is common in qualitative research, where using a 10-25\% subset for IRR is considered typical for ensuring the reliability of the coding scheme \cite{o2020intercoder}. This initial human-human check yielded an inter-coder reliability (IRR) with a Cohen's Kappa \cite{cohen1960coefficient} of 0.7798, indicating substantial agreement \cite{landis1977measurement}. After resolving disagreements through discussion, we finalized the training ground truth. The initial Kappa score before this process was 0.1765, highlighting the necessity for refinement. We then refined the prompt over six iterations. The changes primarily involved adding more explicit examples for each category, clarifying boundaries between easily confused categories (e.g., \textit{Training Infrastructure} vs. \textit{Parameter Tuning}), and linking definitions to specific file types and configuration keys. This iterative process concluded when the LLM achieved a stable and high agreement with a Kappa score of 0.9068.

    \item  \textbf{Final Prompt Testing Phase (Testing):} To validate the generalizability of the refined prompt and ensure it was not overfitted, we tested it on an independent, held-out test set. This set consisted of 384 randomly sampled commits, a size calculated using a standard formula to be statistically significant for our population of over 680,000 commits (providing a 95\% confidence level with a 5\% margin of error). The ground truth for this test set was established using the same rigorous two-author process. First, to confirm the quality of our manual labels, we measured the IRR between the two human coders on a subset of 40 commits (approx. 10\% of the test set), achieving a Cohen's Kappa of 0.8150. This score indicates "almost perfect" agreement and confirms the robustness of our ground truth. Having validated our manual labels, we then evaluated the final prompt against the entire 384-commit test set. The LLM achieved a Cohen's Kappa score of 0.8568, demonstrating a high degree of strong and generalizable performance. This robust result gave us confidence to use the validated prompt for classifying the remainder of the dataset.
\end{enumerate}

\begin{figure}[h!]
\centering
\includegraphics[width=\textwidth]{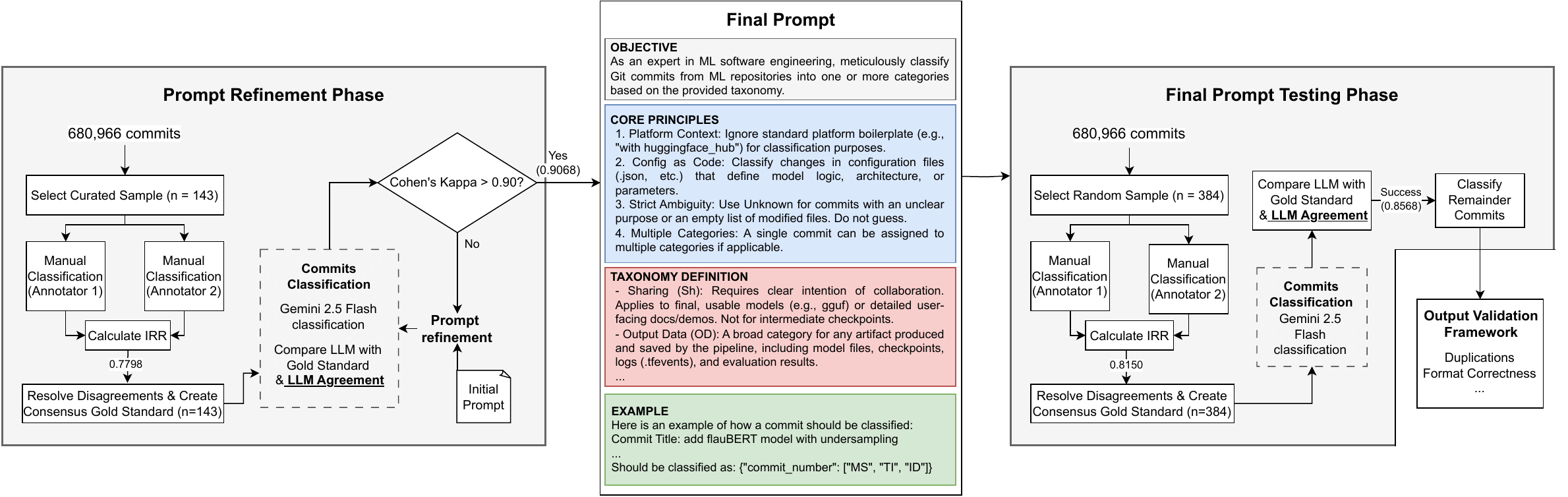}
\caption{Process of refining the LLM prompt through consecutive evaluation rounds and final validation.}
\label{fig:LLM_Validation}
\Description{Process of refining the LLM prompt through consecutive evaluation rounds and final validation.}
\end{figure}

After the LLM classification, we performed a final \textit{automated} validation on all results using a framework implemented as part of our pipeline. This step checked for programmatic errors (e.g., malformed JSON) or duplications, ensuring the final dataset was reliable. This multi-stage validation process is illustrated in Fig.~\ref{fig:LLM_Validation}.

\paragraph{{\textbf{Enhanced Dataset for BN Analysis:}}}

To accommodate the commit information in the BN for RQ2 and RQ3, we create another dataset where the commits are in tabular form, ordered by commit order, with the commit type variable one-hot encoded. Moreover, to improve the BN fitting to our commit changes and how they evolve, we expand our commits dataset with additional variables that could be complementary to the class variables (the commit types). Some of the added attributes are:

\begin{itemize}
\item \textbf{time\_since\_model\_creation}: Time elapsed since the creation of the model in seconds. This helps evaluate if the project's maturity affects the types of commits.
\item \textbf{time\_between\_commits}: Time difference between the current commit and the previous one in seconds. This helps evaluate if the types of commits depend on their frequency.
\item \textbf{commit\_size\_change}: Change in the commit size compared to the previous commit. To evaluate fluctuations in commit sizes.
\item \textbf{collaboration\_intensity:} A measure of the number of unique contributors and the frequency of their contributions, indicating the level of collaborative activity on the project.
\end{itemize}

Moreover, to facilitate the analysis using BNs, several continuous variables are discretized into binary categories as follows:

\begin{itemize}
    \item \textbf{popularity\_high}: This binary variable is set to \texttt{True} if the model's popularity exceeds the 95th percentile and \texttt{False} otherwise. We focus on the 95th percentile for several reasons. Firstly, examining extreme groups is a common practice in empirical studies to understand distinct characteristics \cite{osborne2004power}. Secondly, our popularity metric, calculated based on factors like downloads and likes (details in \cite{castano2023analyzing}), exhibits a highly right-skewed distribution across the 99,784 models with available data: the mean popularity was 0.0004, while the median (50th percentile) and even the 75th and 90th percentiles were 0.0000. The 95th percentile corresponded to a popularity score of 0.0004. This cutoff effectively isolates the top ~5\% of models (4,983 models or 4.99\% in our dataset had popularity > 0.0004), representing a distinct group with significantly higher visibility and allowing for a clearer analysis of patterns associated with high adoption.
    \item \textbf{commit\_size\_category\_high}: This variable is set to \texttt{True} if the \texttt{commit\_size\_category} is either "large" or "very\_large", and \texttt{False} if it is "small". The categorization into "small", "large", and "very\_large" is created using quartiles from the dataset.
    \item \textbf{collaboration\_intensity\_high}: This binary variable is set to \texttt{True} if \texttt{collaboration\_\allowbreak intensity} exceeds 2, and \texttt{False} otherwise. We choose this threshold because most values are 2 or less, so values greater than 2 represent higher collaboration intensity.

\end{itemize}

These discretized variables enable the BNs to effectively model the relationships and dependencies between different commit types and project characteristics. Additionally, the dataset was enriched with a variety of further attributes. Further details on the computation behind these attributes and the structure of this dataset can be found in the replication package \cite{zenodo}.

\subsubsection{\textbf{Dataset Splits}}

To address our RQs, we end up with four distinct datasets:

\begin{enumerate}
    \item \textbf{Commits from 100,000 models}: This dataset, containing over 680,000 classified commits randomly sampled from the HF population up to May 2025, is used for the study of RQ1. It provides a large-scale overview of commit types and patterns across a diverse range of models.
    
    \item \textbf{Commits from 14,343 models (with >= 10 commits)}: This dataset focuses on all models from our initial 100k sample that have at least 10 commits, enabling a more detailed evolutionary study for RQ2. It allows for the analysis of patterns and dependencies in commit sequences over time.

    \item \textbf{Releases from 202 models}: This dataset includes models identified through a filtering process (including having at least 5 and at most 50 plausible version tags), totaling 2,251 releases. It supports the analysis for RQ3.1, RQ3.2, and RQ3.3, facilitating the investigation of release patterns and their evolution across different models using repository-level data.

    \item \textbf{Metadata of releases from 28 models}: This dataset focuses on the detailed internal metadata extracted from the releases of a subset of the 202 models above, selected for parseable file formats. It is specifically used for RQ3.4 to analyze the evolution of model parameters and configurations over time by parsing the model files.
\end{enumerate}

Each dataset is designed to capture a specific aspect of model evolution, enabling a focused and thorough examination of the distinct facets addressed by our research questions.

\subsection{Data Analysis}

In this section, we describe our approach for analyzing the data to answer our RQs. We aim to provide a clear and reproducible account of how we analyzed the data and derived conclusions. 

\subsubsection{Initial Analysis of File Changes in ML Repositories}

Before addressing our primary research questions, we conduct a preliminary analysis to understand the typical composition and common change patterns of files within HF model repositories. This contextual analysis is performed on our randomly sampled dataset of 100,000 models (and their associated commits, totaling over 1 million) collected up to May 2025. The goal was to provide background on the artifacts commonly found and modified in these repositories.

Our methodology involves several steps, detailed further in the replication package \cite{zenodo}:
\begin{itemize}
    \item \textbf{Identifying Common Files and Model Extensions:} We aggregate filenames across all commits in the 100k sample to identify the most frequently occurring files overall. We also develop a heuristic (based on commit frequency and a predefined priority list of common extensions) to determine the primary 'model file' extension (e.g., `.safetensors`, `.bin`) for each repository and analyze the distribution of these extensions.
    \item \textbf{Analyzing Co-occurrence Patterns:} To understand which files are commonly modified together within the same commit (suggesting related tasks), we perform a co-occurrence analysis. For each commit, we identified pairs of files that are changed simultaneously. We aggregate these pairs across commits for each model.
    \item \textbf{Evolutionary Co-occurrence Analysis:} To explore if co-occurrence patterns change over a project's lifecycle, we divide each model's commit history into five equal stages based on commit count (0-20\%, 20-40\%, ..., 80-100\%). This division into five stages is adopted for consistency with our prior lifecycle analysis framework presented in \cite{castano2023analyzing}, which also utilized five stages to delineate distinct phases of model evolution, allowing us to potentially correlate findings across studies. We then aggregate the file co-occurrence frequencies separately for each stage across models. We construct a co-occurrence graph for each stage (nodes=files, edges=weighted by co-occurrence frequency) and apply the Louvain community detection algorithm \cite{blondel2008fast} to identify clusters of files frequently edited together at different lifecycle stages. We specifically analyze the co-occurrence patterns involving the identified primary model file.
\end{itemize}
The results of this analysis, detailed in Section \ref{sec:results_file_changes}, provide essential context on the typical file structures and modification workflows observed in the sampled HF repositories.

\subsubsection{\textbf{RQ1 Analysis}}
To address RQ1.1, we construct time-series graphs to visualize the evolution of commit types across the HF ecosystem. We calculate the proportion of each commit type on a quarterly basis and plot these proportions over time. This analysis helps us identify trends in the relative frequency of different commit types and how they have changed over the lifespan of the platform. Additionally, we learn a BN from the data, including the variables for the commit type and the project phase. A Hill-Climbing algorithm with the \textit{Bayesian Information Criterion (BIC)} score is applied to learn the structure and the parameters are estimated by maximum likelihood  ~\cite{scanagatta2019survey}. For large samples, like the ones used in this paper, this is equivalent to assuming a Dirichlet-Multinomial conjugate model and using MAP (maximum a posteriori) estimators of the parameters~\cite[Ch.3.4]{murphy}. The convergence of the MAP estimator to the maximum likelihood estimator holds for any choice of the parameters of the Dirichlet prior, but in particular, if the prior is selected to be uniform, the MAP estimator is equal to the maximum likelihood. Hence, by using maximum likelihood estimators here, we are assuming uniform priors on the parameters. The analysis of the commit type evolution is performed by calculating  the probability of each phase in the model when observing that a commit is of a certain type.   

For RQ1.2, we employ a multi-faceted approach to examine how project characteristics influence commit types throughout a model's lifecycle:

\begin{itemize}
    \item We calculate correlations between various project characteristics (e.g., model size, time since creation, collaboration intensity, popularity) and the frequency of different commit types. We visualize these correlations using a heatmap to identify strong relationships.
    \item We analyze the distribution of commit types across different project phases (Initial, Early, Mid, Late) based on the number of commits. This helps us understand how commit patterns evolve as projects mature.
    \item We examine how commit types vary across different model size categories (Small, Medium, Large, Very Large), which have been defined based on quartiles, to understand the impact of project scale on development practices.
    \item We investigate the relationship between commit types and the time between commits. Based on the distribution of 680,966 inter-commit times, which is also highly right-skewed (mean: $\approx$158k sec, median: 92 sec), we categorize these intervals into distinct groups that aim to reflect plausible development cadences and data-observed clusters:
        \begin{itemize}
            \item Less than 1 hour: Captures the vast majority (87.8\%) of inter-commit times, representing very rapid, potentially automated, or near-continuous development activity.
            \item 1 hour to 1 day: Represents a significant portion (8.0\%) of commits that might occur within a single focused work session or a typical workday cycle.
            \item 1 day to 1 week: Accounts for 2.1\% of intervals, potentially reflecting work done on a daily or multi-day basis, allowing for overnight reflection or batching of changes.
            \item Greater than 1 week: Also 2.1\%, this category captures commits separated by longer periods, which could signify work on larger features, interruptions, or less frequent maintenance updates.
        \end{itemize}
        These data-informed categories allow for an analysis of how commit type frequencies vary across these different temporal patterns of development activity.
    \item For categorical variables like domain, we create heatmaps showing the distribution of commit types across different categories.
    \item We investigate which commit types are likely to occur simultaneously. For this, we consider the same BN used in RQ1.1. For each possible pair of commit types, we calculate the probability that a commit is of a certain type given that we observe it is of another type.

\end{itemize}

To visualize our findings, we use a combination of line plots for time series data, heatmaps for correlation analyses and categorical distributions, bar plots for comparing averages across categories, and scatter plots for examining relationships between continuous variables.


\subsubsection{\textbf{RQ2 Analysis}} \label{rq2_analysis}

In RQ2, we investigate the patterns in commit sequences. We learn a 2-time step DBN~\cite{murphy2002dynamic}, which essentially serves as a BN for time series analysis by duplicating the variables for two consecutive time steps. These two consecutive time steps ($t_0, t_1$) serve as two waves for our longitudinal study. The variables considered include those related to commit types, as well as other characteristics such the time between commits, the commit size, the collaboration intensity and popularity of the project. The same learning algorithms used in the static case have been applied here, but with the constraint that arcs cannot go from a variable to its counterpart in the previous time step. 

To analyze the dependencies between different commit types over time (RQ2.1), we calculate the probability of each commit type in two consecutive time steps. Subsequently, we run the same queries while conditioning on the aforementioned characteristics to address RQ2.2.

\subsubsection{\textbf{RQ3 Analysis}}

Our goal is to replicate the analytical approaches used for RQ1 and RQ2 on the releases of models instead of commits, leveraging the dataset of 2,251 releases from 202 models. Additionally, for RQ3.4, we analyze the metadata differences of releases from a subset of 28 models.

To address RQ3.1, we replicate the analysis methodology used in RQ1.1. This involves constructing time-series graphs to visualize the distribution and evolution of release types over time. We calculate the proportion of each release type on a quarterly basis and plot these proportions to identify trends and patterns in release activities.

For RQ3.2, we employ the same analytical methods as in RQ1.2 to examine how project characteristics influence release types. This includes calculating correlations between various project characteristics (e.g., model size, time since creation, collaboration intensity) and the frequency of different release types. We analyze the distribution of release types across different project phases (Initial, Early, Mid, Late) to understand how release patterns change as projects mature. Additionally, we examine how release types vary across different model size categories (Small, Medium, Large, Very Large) to determine the impact of project scale on release practices. We also investigate the relationship between release types and the time between releases, using the same time interval categories as defined for commits ($<$1 hour, 1 hour - 1 day, 1 day - 1 week, $>$1 week) to allow for comparable analysis of temporal patterns in release activities.

For RQ3.3, we use the same methodology as in RQ2.1 to analyze the patterns in release sequences using the 202-model dataset. We learn a 2-time step DBN to model the dependencies between different release types over time. This involves calculating the probability of each release type in two consecutive time steps to identify temporal dependencies. Additionally, we run queries to determine how these dependencies are influenced by other characteristics such as time between releases, release size, collaboration intensity, and model popularity.

For RQ3.4, we focus on the detailed metadata differences of releases from the subset of 28 models. This involves extracting and analyzing the differences in metadata keys (e.g., parameter names, tensor shapes, data types, total parameters, optimizer states, mean values, nested structures) between successive releases by parsing the model files. We calculate the distribution of these differences to understand how model parameters and configurations change over time.

\section{RESULTS} \label{sec:results}

This section is structured as follows: we begin with an initial analysis of the organization of files in ML repositories on HF and how they change in commits. This is followed by the results addressing our three research questions: RQ1, RQ2, and RQ3.

\subsection{File changes in ML Repositories} \label{sec:results_file_changes} 

Understanding the typical organization and common file changes in ML repositories is crucial for contextualizing our analysis of model evolution on the HF platform. In this section, we provide an overview of how models are structured within repositories on HF and discuss general patterns observed in file changes. For this preliminary, contextual analysis, we examine the full set of approximately 1 million commits from our initial sample of 100,000 models to provide the broadest possible overview of the repository landscape. It is important to note that for the main statistical analyses in the subsequent sections (RQ1, RQ2, and RQ3), this dataset is filtered to 680,000 commits by removing models with more than 500 commits. This step was taken to exclude potential outliers, such as automated repositories, and focus on more typical development patterns, as further detailed in Section~\ref{sec:methodology}.

\subsubsection{Organization of HF ML Repositories}

Repositories on HF for models typically contain various files that collectively define the model, its configuration, and associated metadata. The core components observed frequently in our sample include:

\begin{itemize}
    \item \textbf{Model Files:} Files containing the learned parameters. Based on identifying the most frequently committed model file extension per repository in our sample, the most common primary model file extensions were \texttt{.safetensors} (approx. 33.6k models), \texttt{.bin} (approx. 28.9k models), and \texttt{.gguf} (approx. 6.5k models), followed by formats like \texttt{.zip}, \texttt{.pth}, \texttt{.pt}, and \texttt{.pkl}. The rise of \texttt{.safetensors} reflects a community shift towards this safer format.
    \item \textbf{Configuration Files:} JSON files like \texttt{config.json}, \texttt{tokenizer\_config.json}, and \texttt{genera} \texttt{tion\_config.json} defining model architecture, tokenizer settings, etc., appear frequently.
    \item \textbf{Tokenizer Files:} Files like \texttt{tokenizer.json}, \texttt{vocab.json}, \texttt{merges.txt}, or \texttt{tokenizer.mod} \texttt{el} specifying text processing are common, especially for NLP models.
    \item \textbf{Metadata and Documentation:} \texttt{README.md} for documentation and \texttt{.gitattributes} for repository management are ubiquitous.
\end{itemize}

\subsubsection{General Metrics on File Changes}

To provide context on how files change, we analyzed the frequency of edits involving different filenames across the commits in our sample. The most commonly edited filenames were:

\begin{itemize}
    \item \texttt{.gitattributes} (edited in commits across $\sim$98.8k repositories in the sample): Reflects its importance for Git Large File Storage (LFS) configuration, essential for large model files.
    \item \texttt{README.md} ($\sim$67.9k repos): Indicates frequent updates to documentation.
    \item Tokenizer-related files: \texttt{tokenizer\_config.json} ($\sim$43.7k repos), \texttt{special\_tokens\_map.jso} \texttt{n} ($\sim$42.9k repos), \texttt{tokenizer.json} ($\sim$38.6k repos) show frequent modifications, highlighting the importance of input processing.
    \item \texttt{config.json} ($\sim$42.5k repos): Common edits reflect changes to model architecture or parameters.
    \item Training-related files like \texttt{training\_args.bin} ($\sim$25.3k repos) and adapter files (\texttt{adapter\_con} \texttt{fig.json}, \texttt{adapter\_model.safetensors}) also appeared in the top 10 most frequently edited filenames, indicating common patterns related to training setups and model adaptation (e.g., LoRA).
\end{itemize}
(Counts reflect presence in commits across the unique repositories in the 100k sample, exact counts in replication package \cite{zenodo}).

\subsubsection{Patterns in File Changes and Their Evolution}

\begin{figure*}[ht] 
    \centering
    \begin{subfigure}[b]{0.48\textwidth} 
        \centering
        \includegraphics[width=\textwidth]{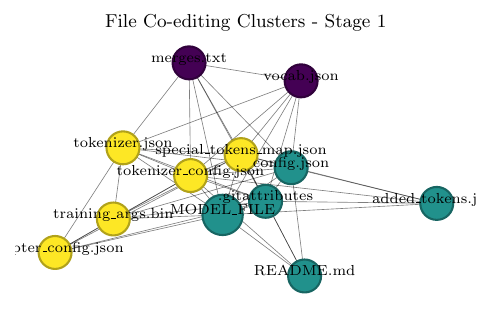} 
        \caption{Stage 1 Clusters}
        \label{fig:stage1_clusters}
    \end{subfigure}
    \hfill 
    \begin{subfigure}[b]{0.48\textwidth} 
        \centering
        \includegraphics[width=\textwidth]{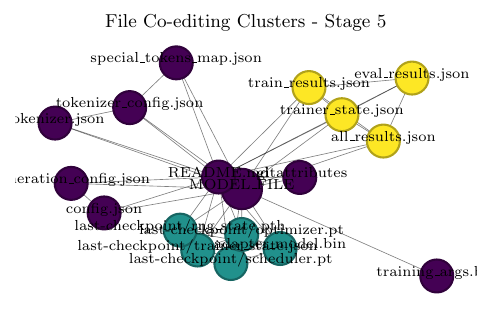} 
        \caption{Stage 5 Clusters}
        \label{fig:stage5_clusters}
    \end{subfigure}
    \caption{Evolution of file co-editing clusters across lifecycle stages (based on Louvain community detection on file co-occurrence graph within commits). Stage 1 (left) shows distinct clusters for core model/config files and tokenizer files. Stage 5 (right) shows different groupings, with README.md strongly linked to the model file.}
    \label{fig:file_clusters} 
\end{figure*}

To understand common workflows, we analyzed which files are frequently edited together within the same commit, based on systematic frequency counting across commits in our 100k sample. We also explored how these co-occurrence patterns evolve across five lifecycle stages (approximated by commit count quartiles). Key observations include:

\begin{itemize}
    \item \textbf{Central Role of the Model File:} Across all lifecycle stages, the primary model file (identified per repository, e.g., `.safetensors`, `.bin`, represented generically as `MODEL\_FILE` in our analysis) exhibits high connectivity in the co-occurrence graph (degree centrality > 0.95 in all stages). This confirms its central role, as changes to it often coincide with other file modifications.
    \item \textbf{Common Co-occurrence Pairs:}
        \textit{Early Stage (Stage 1):} The model file is most frequently co-edited with repository setup files (\texttt{.gitattributes}), core configuration (\texttt{config.json}), and tokenizer files (\texttt{tokenizer\_config.json}, \texttt{special\_tokens\_map.json}, \texttt{tokenizer.json}). This reflects initial model definition and input processing setup. Tokenizer files also form their own distinct co-occurrence cluster in this stage.
        \textit{Mid-Stages (Stages 2-4):} Co-edits between the model file and checkpoint/training state files (e.g., files within `last-checkpoint/`, `training\_args.bin`) become more prominent, alongside continued co-edits with \texttt{.gitattribu} \texttt{tes} and \texttt{README.md}. This suggests a focus on the iterative training and saving process. Distinct clusters emerge for tokenizer components and sometimes framework-specific files.
        \textit{Late Stage (Stage 5):} The model file is most strongly co-edited with documentation (\texttt{README.md}) and repository management (\texttt{.gitattributes}). Co-edits with checkpoint files persist but might be relatively less dominant compared to README co-edits than in mid-stages. Configuration and tokenizer files often form their own clusters distinct from the primary model/README group. The contrasting structures of Stage 1 and Stage 5 are illustrated in Figure~\ref{fig:file_clusters}.
    \item \textbf{Decreasing Model File Edit Proportion:} Interestingly, the proportion of total file edits within commits that involve the primary model file decreases significantly in the last stage (Stage 5: ~19\%) compared to earlier stages (Stages 1-4: ~34-37\%). This might indicate a shift towards documentation, final configuration, or other activities relative to direct model weight updates as a project matures or nears a release state.
\end{itemize}
These patterns, derived quantitatively from commit histories, suggest common workflows: initial setup involves model, config, and tokenizer files; mid-lifecycle focuses on iterative training involving model and checkpoint files; and later stages emphasize documentation alongside model updates, with potentially fewer direct model weight modifications relative to other activities. The clustering analysis (details and visualizations in replication package \cite{zenodo}) provides further evidence for these groupings of related files frequently modified together throughout the development lifecycle.

\subsection{RQ1: Patterns in Commit Types and Their Evolution}

\subsubsection{Distribution and Evolution of Commit Types across the HF Ecosystem (RQ1.1)}

To understand the landscape of changes on models, we analyzed the distribution and evolution of commit types across the ecosystem.
\paragraph{\textbf{Distribution of Commit Types:}}

The analysis of over 680,000 commits reveals that \textit{Output Data} (422,190 commits), \textit{Project Metadata} (241,090 commits), and \textit{Sharing} (226,504 commits) are the three most frequent types of commits. This suggests that developers on the platform predominantly focus on managing the data artifacts produced by models, handling repository-level metadata, and activities related to sharing and collaboration.

Conversely, \textit{Remove Dependency} (373 commits), \textit{Update Dependency} (2,268 commits), and \textit{Input Data} (6,562 commits) are the least common commit types. The low frequency of dependency-related changes suggests that these are infrequent, isolated events in the lifecycle of most models on HF. The rarity of \textit{Input Data} commits may indicate that raw data sources are established early and change infrequently, with more activity focused on pre-processing and model execution.

\paragraph{\textbf{Evolution of Commit Types Over Time:}}

Fig.~\ref{fig:evolution_over_time} illustrates the evolution of commit types over time, with the analysis extending to Q1 2025. The graph reveals several important trends:

\begin{figure}[h]
\centering
\includegraphics[width=0.55\textwidth]{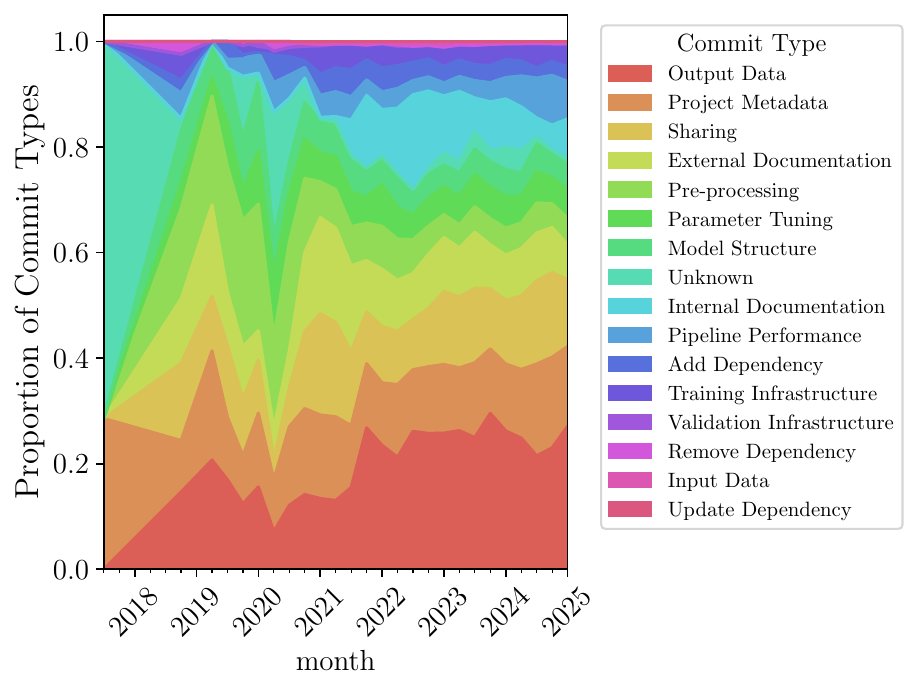}
\caption{Evolution of Commit Types Over Time}
\label{fig:evolution_over_time}
\end{figure}

\begin{itemize}
\item \textit{Output Data} commits have shown a significant and sustained increase in proportion, especially from 2022 onwards, becoming the most dominant commit type in recent quarters (e.g., 0.295 in 2023Q4, 0.271 in 2025Q1). This highlights a growing emphasis on managing, storing, and versioning the products of model execution.
\item \textit{Project Metadata} commits were consistently high in the early years and have remained a very significant commit type, indicating a continued focus on managing repository-level configurations and identifiers (e.g., 0.122 in 2023Q4, 0.151 in 2025Q1).
\item \textit{Sharing} related commits have markedly increased in proportion since late 2022, underscoring the collaborative nature and dissemination efforts on the platform (e.g., peaking at 0.160 in 2024Q4).
\item \textit{Internal Documentation} (which includes logging artifacts) has also seen a noticeable rise in proportion, particularly from 2021 through 2023, suggesting an increased focus on tracking the internal state of development and training processes.
\item \textit{External Documentation} commits peaked in proportion in early 2021 but have since stabilized at a lower, yet still significant, level.
\item In contrast, \textit{Training Infrastructure} commits have remained a consistently low proportion of all changes throughout the observed history, suggesting that fundamental changes to training logic are less frequent than other activities like managing model outputs or metadata.
\item \textit{Model Structure} and \textit{Parameter Tuning} commits have seen their relative proportions decline over time, possibly indicating a maturation of the ecosystem toward using and refining existing architectures rather than constantly creating new ones from scratch.
\end{itemize}

The pronounced rise of \textit{Output Data}, \textit{Sharing}, and \textit{Internal Documentation} commits suggests an evolving ecosystem focused on the operational aspects of ML: managing model artifacts, facilitating their use, and tracking the experimental process. The relative decline in core architectural and tuning commits may point towards a trend of fine-tuning and adapting powerful base models, a practice widely observed in the field \cite{alvarez2024impact}.

\paragraph{\textbf{Evolution of Commit Types Over Project Lifecycle: }}

\begin{figure}[h]
\centering
\includegraphics[width=0.75\textwidth]{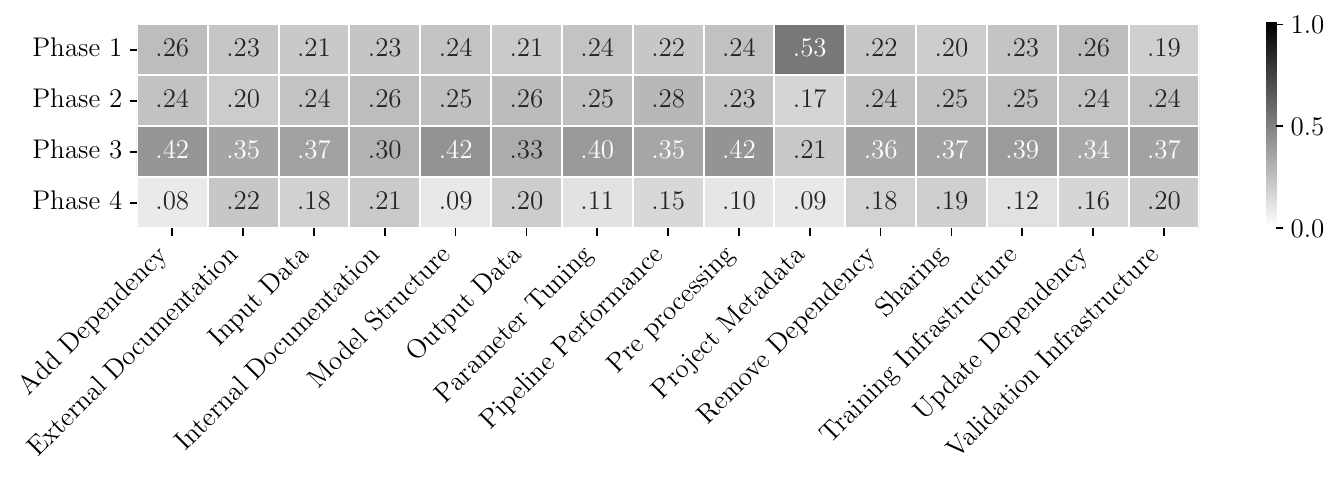}
\caption{Evolution of Commit Types Over a Project Lifecycle}
\label{fig:commit_types}
\Description{Evolution of Commit Types Over a Project Lifecycle}
\end{figure}

Fig. \ref{fig:commit_types} allows for the analysis of the distribution of each commit type across the different phases of the project. Specifically, it shows, calculated from the BN, the probability $P(phase|C=1)$ where $C$ is the one-hot encoded variable representing each commit type. That is, the probability of each phase given that a particular type of commit occurs. From these probabilities, we observe that \textit{Project Metadata} commits are predominantly found in the first phase (quartile 1 of project lifecycle based on commit count), with a probability of 0.528. This highlights the importance of initial setup and configuration at the beginning of a project.

The middle of the lifecycle emerges as a period of intense, diverse activity. Phase 3 (the third quartile) shows the highest probability for a wide range of commit types, including \textit{Add Dependency} (0.417), \textit{Model Structure} (0.424), \textit{Pre-processing} (0.422), and \textit{Training Infrastructure} (0.392). This suggests that the core iterative development, adding dependencies, altering model architecture, refining data preparation, and adjusting training logic, is concentrated in this mature development stage.

\textit{Parameter Tuning} commits are also most dominant in Phase 3 (0.401), suggesting that fine-tuning efforts are concentrated in these later development stages. Other types like \textit{Output Data} (0.327) and \textit{Sharing} (0.367) also peak in Phase 3, reflecting the generation of artifacts and sharing of results during this active development period. This concentration of diverse, core ML tasks in the mid-to-late lifecycle underscores the iterative refinement process central to model development.

\begin{figure}[h]
\centering
\includegraphics[width=0.75\textwidth]{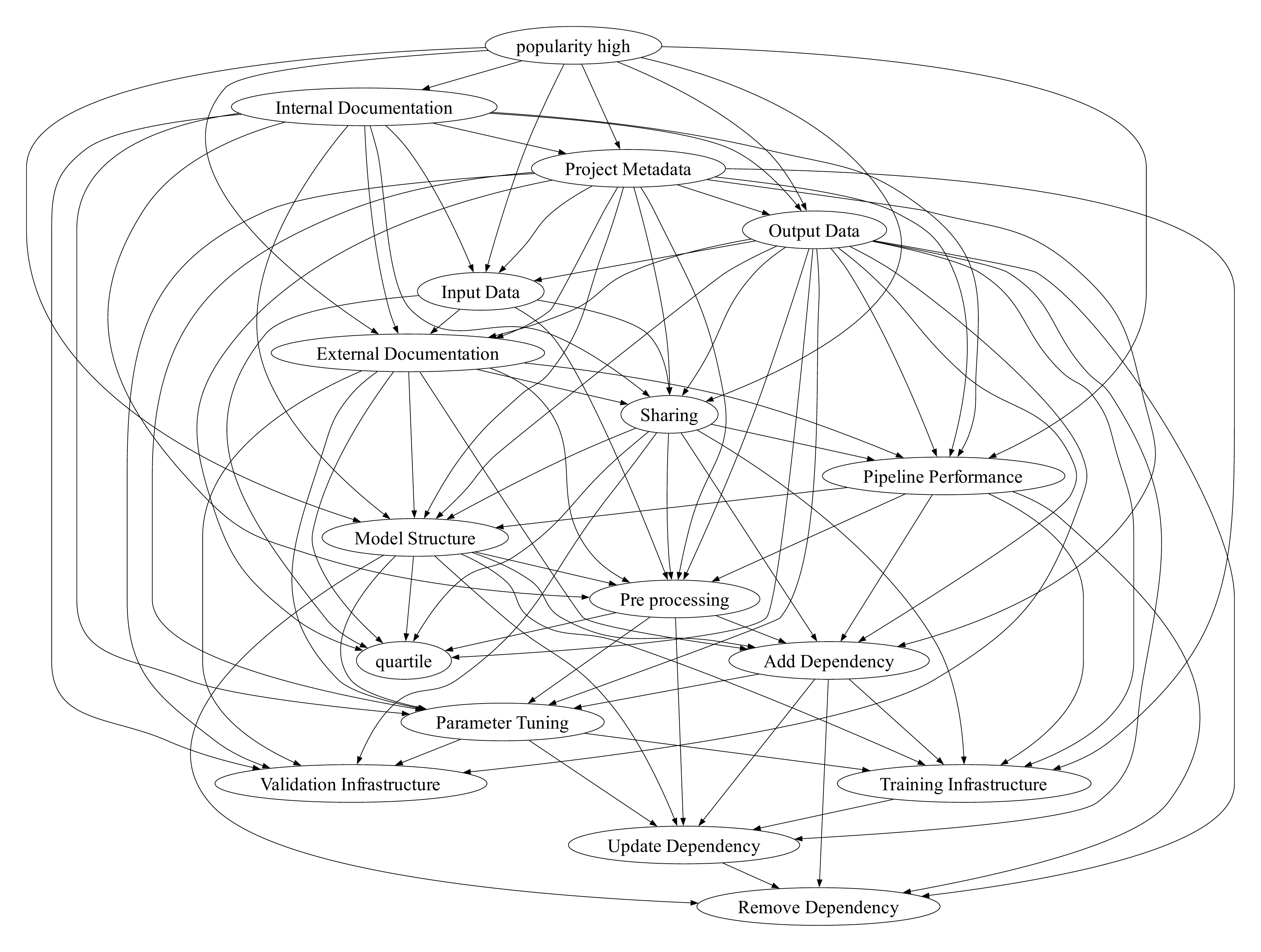}
\caption{Structure of the trained BN used for commit analysis}
\label{fig:commits_static_time}
\end{figure}

The structure of the trained BN used in this analysis can be seen in Fig. \ref{fig:commits_static_time}.

\subsubsection{Association of Project Characteristics with Commit Types (RQ1.2)}
Our analysis of how project characteristics influence commit types throughout a model's lifecycle revealed several interesting patterns.
\paragraph{\textbf{Correlations between Project Characteristics and Commit Types: }}

\begin{figure}[h]
\centering
\includegraphics[width=0.6\textwidth]{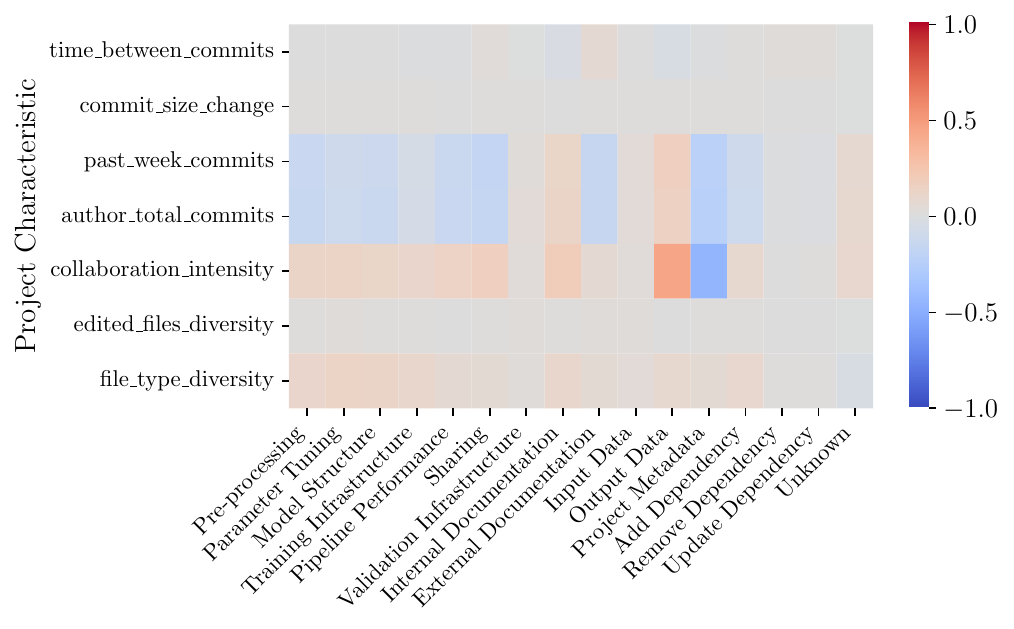}
\caption{Correlations between Project Characteristics and Commit Types}
\label{fig:correlations}
\Description{Correlations between Project Characteristics and Commit Types}
\end{figure}

Fig. \ref{fig:correlations} presents the correlations between project characteristics and commit types. The analysis shows a strong positive correlation between \textit{collaboration\_intensity} and \textit{Output Data} commits (0.440), as well as with \textit{Internal Documentation} (0.175) and \textit{Sharing} (0.163). This suggests that more collaboratively active projects generate more model artifacts and place a higher emphasis on logging and sharing activities, likely to coordinate efforts. Conversely, \textit{collaboration\_intensity} shows a strong negative correlation with \textit{Project Metadata} updates (-0.462), as do metrics like \textit{past\_week\_commits} (-0.227) and \textit{author\_total\_commits} (-0.249). This indicates that as projects become more active and collaborative, the relative focus shifts away from foundational metadata towards the production and management of model-related artifacts and documentation.

\paragraph{\textbf{Commit Types Across Model Sizes: }}

\begin{figure}[h]
\centering
\includegraphics[width=0.6\textwidth]{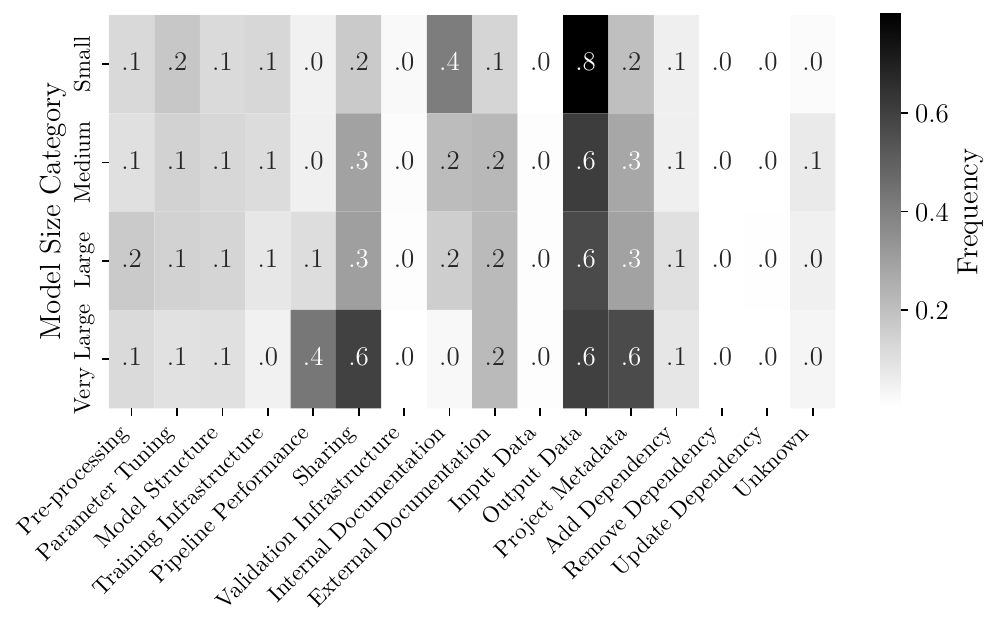}
\caption{Commit Type Distribution by Model Size}
\label{fig:model_size}
\Description{Commit Type Distribution by Model Size}
\end{figure}

The distribution of commit types across different model sizes, as shown in Fig. \ref{fig:model_size}, reveals complex and distinct patterns in development focus, shifting away from a simple linear trend. For Small models, the most dominant activities are overwhelmingly related to the products of the pipeline, with extremely high mean frequencies for \textit{Output Data} (0.800) and \textit{Internal Documentation} (0.407). This suggests that for smaller-scale projects, the core loop is centered on generating and logging results.
As model size increases to Medium and Large, the focus diversifies. While \textit{Output Data} remains very common, activities like \textit{Sharing} and \textit{External Documentation} become more prominent.
For Very Large models, a distinct pattern emerges. \textit{Pipeline Performance} becomes a major focus (mean frequency of 0.426), alongside continued high frequencies for \textit{Sharing} (0.596), \textit{Project Metadata} (0.563), and \textit{Output Data} (0.598). This suggests that at a very large scale, optimizing runtime efficiency and managing the model's public presence and configuration become critical priorities, in addition to generating outputs.

It is important to note that some columns in figures appear as 0 due to rounding, though they are very close to 0 rather than exactly 0. Additionally, the columns do not sum to 1 because commits can belong to multiple types.

\paragraph{\textbf{Commit Types Across ML Domains: }}

\begin{figure}[h]
\centering
\includegraphics[width=0.6\textwidth]{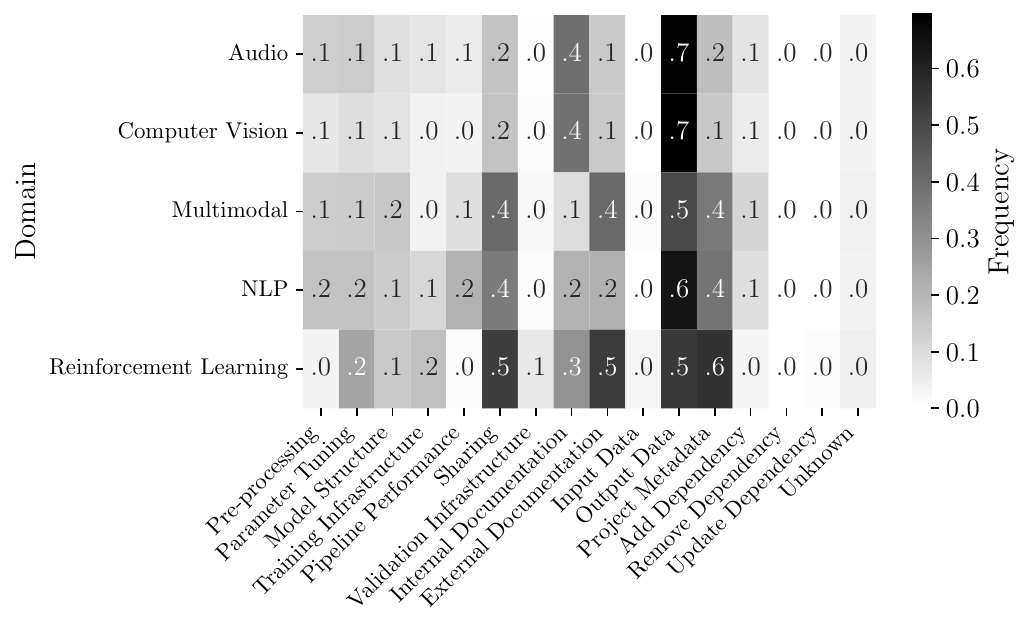}
\caption{Commit Type Distribution Across ML Domains}
\label{fig:ml_domains}
\Description{Commit Type Distribution Across ML Domains}
\end{figure}

The distribution of commit types across various ML domains, illustrated in Fig. \ref{fig:ml_domains}, highlights distinct changes patterns in different areas of ML. Reinforcement Learning (RL) presents a unique profile where collaborative and communicative activities are paramount; it shows the highest mean frequencies for \textit{Sharing} (0.528), \textit{External Documentation} (0.531), \textit{Project Metadata} (0.559), and \textit{Output Data} (0.539). This likely reflects the highly experimental nature of RL, requiring extensive documentation, sharing of results, and management of numerous model artifacts and configurations.
In contrast, NLP models show the highest mean frequency for \textit{Pipeline Performance} (0.210), suggesting a strong focus on optimizing efficiency for this often computationally intensive domain. Domains like Audio and Computer Vision are characterized by a very high frequency of \textit{Internal Documentation} and \textit{Output Data} commits, indicating a development process centered on generating and logging pipeline artifacts.

\paragraph{\textbf{Commit Types by Time Between Commits: }}

\begin{figure}[h]
\centering
\includegraphics[width=0.55\textwidth]{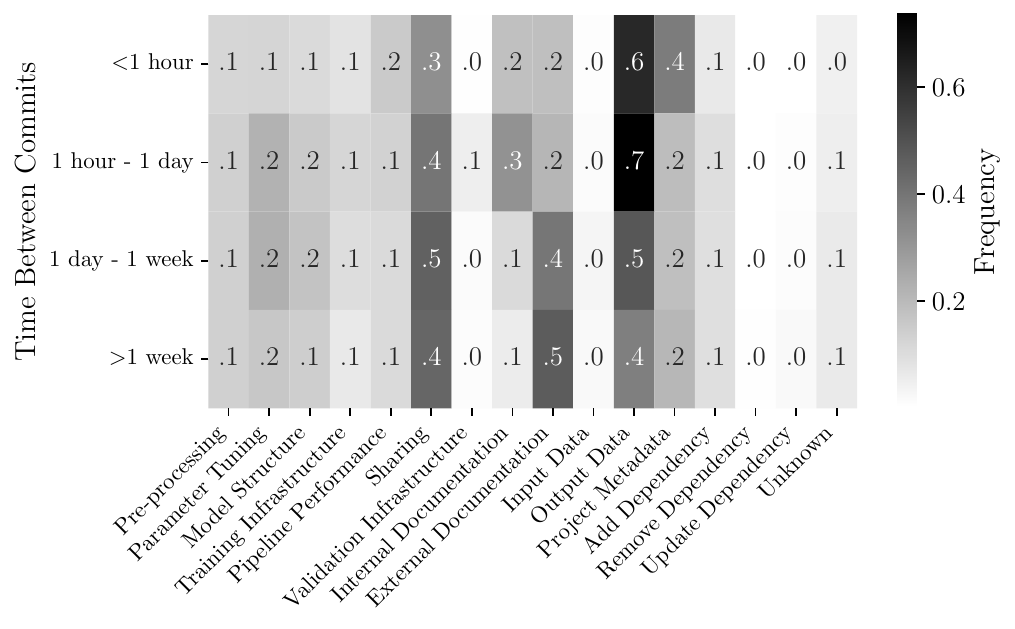}
\caption{Commit Type Distribution by Time Between Commits}
\label{fig:time_between_commits}
\Description{Commit Type Distribution by Time Between Commits}
\end{figure}

The relationship between commit types and the time interval between commits, shown in Fig. \ref{fig:time_between_commits}, reveals interesting temporal patterns in development activities. For very frequent commits (less than 1 hour apart), activities are dominated by the generation of pipeline artifacts, with \textit{Output Data} (mean frequency of 0.619) and \textit{Internal Documentation} (0.183) being most common. Commits related to \textit{Project Metadata} (0.376) are also most frequent in this short interval, suggesting immediate, small updates to project information.
As the time between commits increases, the focus shifts. \textit{External Documentation} shows a clear increasing trend with longer intervals, peaking at a mean frequency of 0.470 for commits made more than a week after the previous one. This pattern strongly indicates that comprehensive documentation updates are often batched or undertaken after longer periods of development. Similarly, \textit{Sharing} and \textit{Parameter Tuning} activities are more frequent when commit intervals are longer (e.g., 1 day to 1 week), suggesting these are more deliberate actions rather than rapid, iterative changes.

\paragraph{\textbf{Dependencies between Commit Types:}}

\begin{figure}[h]
\centering
\includegraphics[width=0.6\textwidth]{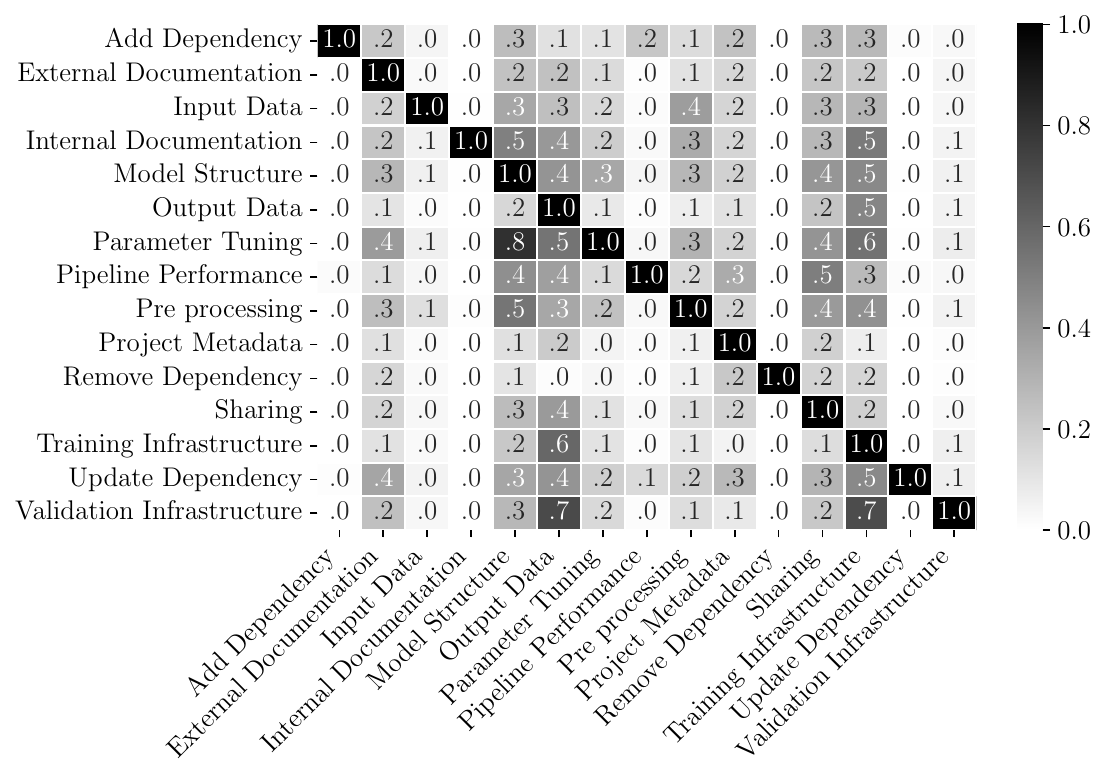}
\caption{Distribution between commit types}
\label{fig:types_prob}
\Description{Distribution between commit types}
\end{figure}

Figure \ref{fig:types_prob} displays the conditional probabilities, derived from the Bayesian Network, of a commit being of a certain type (X) given that it is already classified as another type (Y) (i.e., P(X=1|Y=1)). This analysis reveals several extremely strong co-occurrence patterns, indicating that certain tasks are almost always performed together in a single commit.
\begin{itemize}
    \item Commits involving adding a new dependency (\textit{Add Dependency}) are almost certain to also include changes to \textit{Model Structure} (0.948 probability), \textit{Parameter Tuning} (0.932), \textit{Output Data} (0.905), and \textit{Project Metadata} (0.872). This suggests that adding a core dependency like `transformers` often happens as part of an initial, comprehensive commit that defines the entire model architecture, its parameters, and metadata.
    \item Changes to \textit{Internal Documentation} (such as logging files) have a near-perfect (0.989) probability of co-occurring with \textit{Output Data} commits, confirming that logs and model artifacts are generated and saved in the same pipeline step.
    \item Similarly, a \textit{Pipeline Performance} commit has a 0.963 probability of co-occurring with an \textit{Output Data} commit, likely because performance optimizations (like quantization) result in a new, optimized model file.
    \item The strong link between \textit{Parameter Tuning} and \textit{Model Structure} (0.750) persists, indicating that architectural configurations and hyperparameter settings are frequently modified together.
\end{itemize}
These tight couplings reveal that ML development often proceeds in comprehensive, bundled changes, where a single action (like adding a dependency or optimizing performance) entails modifications across multiple facets of the ML system simultaneously.

\paragraph{\textbf{Commit Types Across Phases and Popularity: }}

\begin{figure}[h]
\centering
\includegraphics[width=0.7\textwidth]{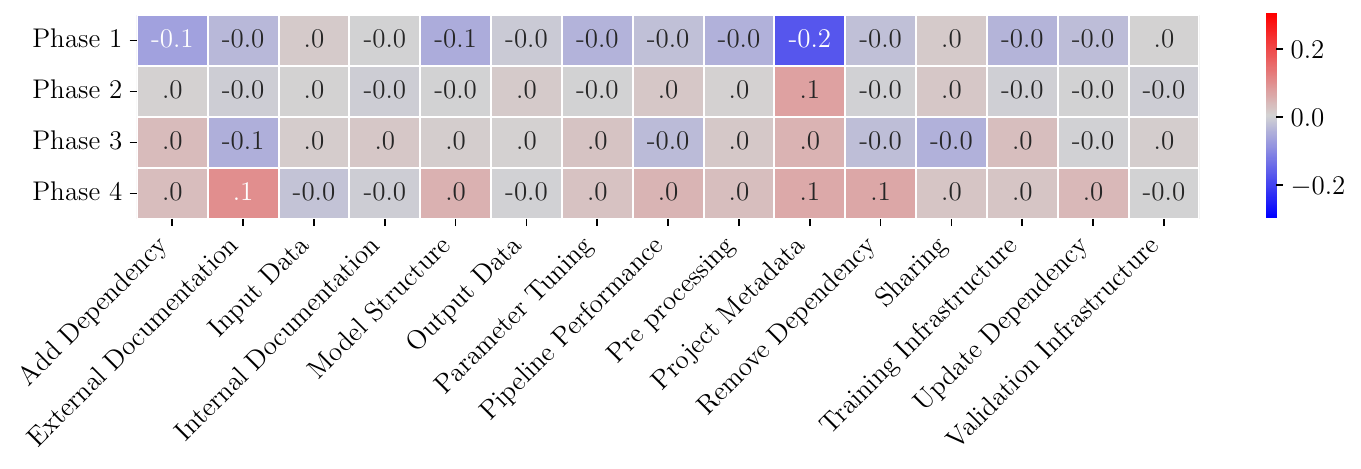}
\caption{Differences in the Commit Distribution between highly popular projects and the rest}
\label{fig:pop_phases}
\Description{Differences in the Commit Distribution between highly popular projects and the rest.}
\end{figure}

Fig. \ref{fig:pop_phases} shows the variation in commit distributions between popular and non-popular projects, specifically the difference between $P(phase|C=1, popular\_high=1)$ and $P(phase|C=1, popular\_high=0)$ where $C$ is each possible type. The analysis reveals that in Phase 1 (the first quartile of commits), highly popular projects exhibit a markedly different pattern. They have a substantially lower proportion of \textit{Project Metadata} commits (difference of -0.176) compared to non-popular projects. Furthermore, they also show a lower likelihood of commits related to \textit{Add Dependency} (-0.071), \textit{Model Structure} (-0.055), \textit{Pre-processing} (-0.048), and \textit{Parameter Tuning} (-0.045) in this initial phase.
Conversely, popular projects show a slightly higher likelihood for \textit{Sharing} commits (+0.012) in Phase 1. This suggests that the early lifecycle of what will become popular projects is less focused on foundational setup and architectural definition within the commit history itself. Instead, they may start with a more complete initial state and have a slightly greater early emphasis on sharing and dissemination.

\begin{myquote}
{\large \textbf{Main Findings for RQ1}}

\vspace{0.2cm}

\textbf{Finding 1}. \textit{\textit{Output Data}, \textit{Project Metadata}, and \textit{Sharing} are the most common commit types, indicating a strong focus on managing model artifacts, repository-level configurations, and collaborative activities in model development on HF.}

\vspace{0.2cm}

\textbf{Finding 2}. \textit{Over time, there has been a significant proportional increase in \textit{Output Data}, \textit{Sharing}, and \textit{Internal Documentation} commits. In contrast, the relative proportions of \textit{Model Structure} and \textit{Parameter Tuning} commits have declined, suggesting a maturing ecosystem focused on artifact management, collaboration, and the refinement of existing architectures.}

\vspace{0.2cm}

\textbf{Finding 3}. \textit{\textit{Project Metadata} commits are predominantly found in the initial phase (Phase 1) of projects. The mid-lifecycle (Phase 3) emerges as the most intense period for core development, showing the highest frequency for \textit{Model Structure}, \textit{Parameter Tuning}, \textit{Pre-processing}, and \textit{Training Infrastructure} commits.}

\vspace{0.2cm}

\textbf{Finding 4}. \textit{Projects with high collaboration intensity are correlated with a higher proportion of \textit{Output Data}, \textit{Internal Documentation}, and \textit{Sharing} commits, and a notably lower proportion of \textit{Project Metadata} commits, suggesting a focus on artifact generation and communication in highly collaborative environments.}

\vspace{0.2cm}

\textbf{Finding 5}. \textit{Development priorities shift significantly with model scale. Small models are dominated by \textit{Output Data} and \textit{Internal Documentation} generation. Very Large models, however, show a much stronger focus on \textit{Pipeline Performance}, \textit{Sharing}, and \textit{Project Metadata}, indicating that efficiency and public-facing management become critical at scale.}

\vspace{0.2cm}

\textbf{Finding 6}. \textit{\textit{Output Data} and \textit{Project Metadata} commits are most frequent in very short intervals (<1 hour), while \textit{External Documentation} and \textit{Sharing} updates predominantly occur after longer periods (over a day or week), indicating rapid iteration on artifacts versus deliberate, batched communication.}

\vspace{0.2cm}

\textbf{Finding 7}. \textit{Popular projects tend to have a lower proportion of \textit{Project Metadata}, \textit{Model Structure}, and \textit{Parameter Tuning} commits early in their lifecycle, suggesting their initial state may be more defined, with a slightly greater early focus on \textit{Sharing}.}

\vspace{0.2cm} 

\textbf{Finding 8}. \textit{Analysis of co-occurring commit types reveals extremely tight bundling of tasks. For instance, adding a new dependency is almost always committed alongside changes to \textit{Model Structure} (0.948 probability) and \textit{Parameter Tuning} (0.932). Similarly, \textit{Internal Documentation} and \textit{Output Data} are nearly perfectly correlated (0.989 probability), highlighting comprehensive, multi-faceted updates within single commits.}
\end{myquote}

\subsection{RQ2: Patterns in the Evolution of Commit Changes}

\subsubsection{Dependencies between Different CommitTypes over Time (RQ2.1)}

\begin{figure}[h]
\centering
\includegraphics[width=0.6\textwidth]{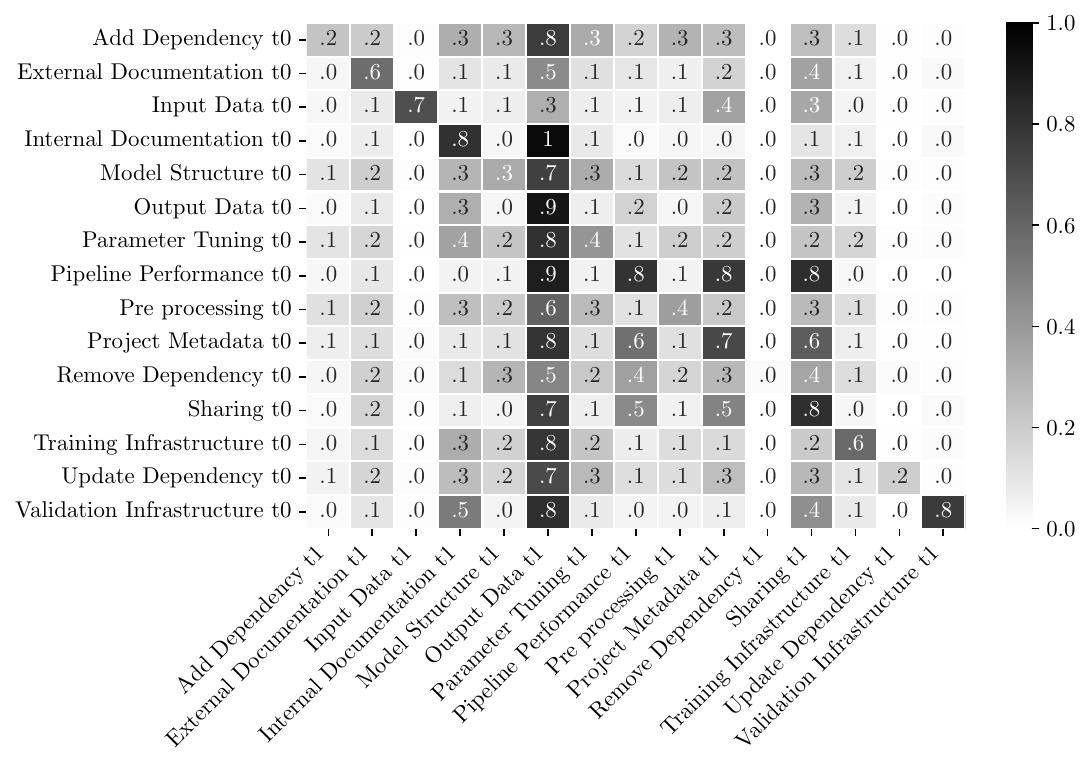}
\caption{Transition probabilities between types in two consecutive commits}
\label{fig:temp_types}
\Description{Transition probabilities between types in two consecutive commits.}
\end{figure}

Fig. \ref{fig:temp_types} shows the probability of each commit type in two consecutive commits computed from the DBN. That is, the probability \( P(C_{t}|C_{t-1}) \) where \( C_{t} \) denotes the commit type at time \( t \). In the context of these figures, \( t0 \) refers to the commit type at the previous time step (\( C_{t-1} \)), and \( t1 \) refers to the commit type at the current time step (\( C_t \)). For example, if \( P(\textit{Input Data at } \allowbreak t1 | \textit{Remove Dependency at } t0) = 0.3 \), this means there is a 30\% probability that an \textit{Input Data} commit occurs immediately after a \textit{Remove Dependency} commit. The DBN analysis reveals strong temporal dependencies, particularly a tendency for activities to be clustered. Many commit types have a high probability of being followed by a commit of the same type. This self-transition is particularly strong for \textit{Output Data} (0.916), \textit{Sharing} (0.815), \textit{Internal Documentation} (0.807), \textit{Pipeline Performance} (0.800), and \textit{Validation Infrastructure} (0.767). This indicates that when developers focus on these areas, they often do so across multiple, consecutive commits.

Beyond self-transitions, the most powerful pattern observed is the prevalence of transitions leading to \textit{Output Data} commits. A commit involving \textit{Internal Documentation} is followed by an \textit{Output Data} commit with 0.961 probability. Similarly, \textit{Pipeline Performance} (0.879), \textit{Parameter Tuning} (0.808), \textit{Model Structure} (0.749), and \textit{Add Dependency} (0.762) commits are all highly likely to be immediately followed by an \textit{Output Data} commit. This strongly suggests a common workflow where any significant change (to code, parameters, performance, or logging) triggers a new pipeline execution, the results of which (the model artifacts, i.e., \textit{Output Data}) are saved in the subsequent commit. Other notable dependencies include performance optimizations leading to sharing activities (\(P(\textit{Sharing}_{t1} | \textit{Pipeline Performance}_{t0}) = 0.818\)) and project metadata changes being followed by performance updates (\(P(\textit{Pipeline Performance}_{t1} | \textit{Project Metadata}_{t0}) = 0.562\)).

\begin{figure}[h]
\centering
\includegraphics[width=0.75\textwidth]{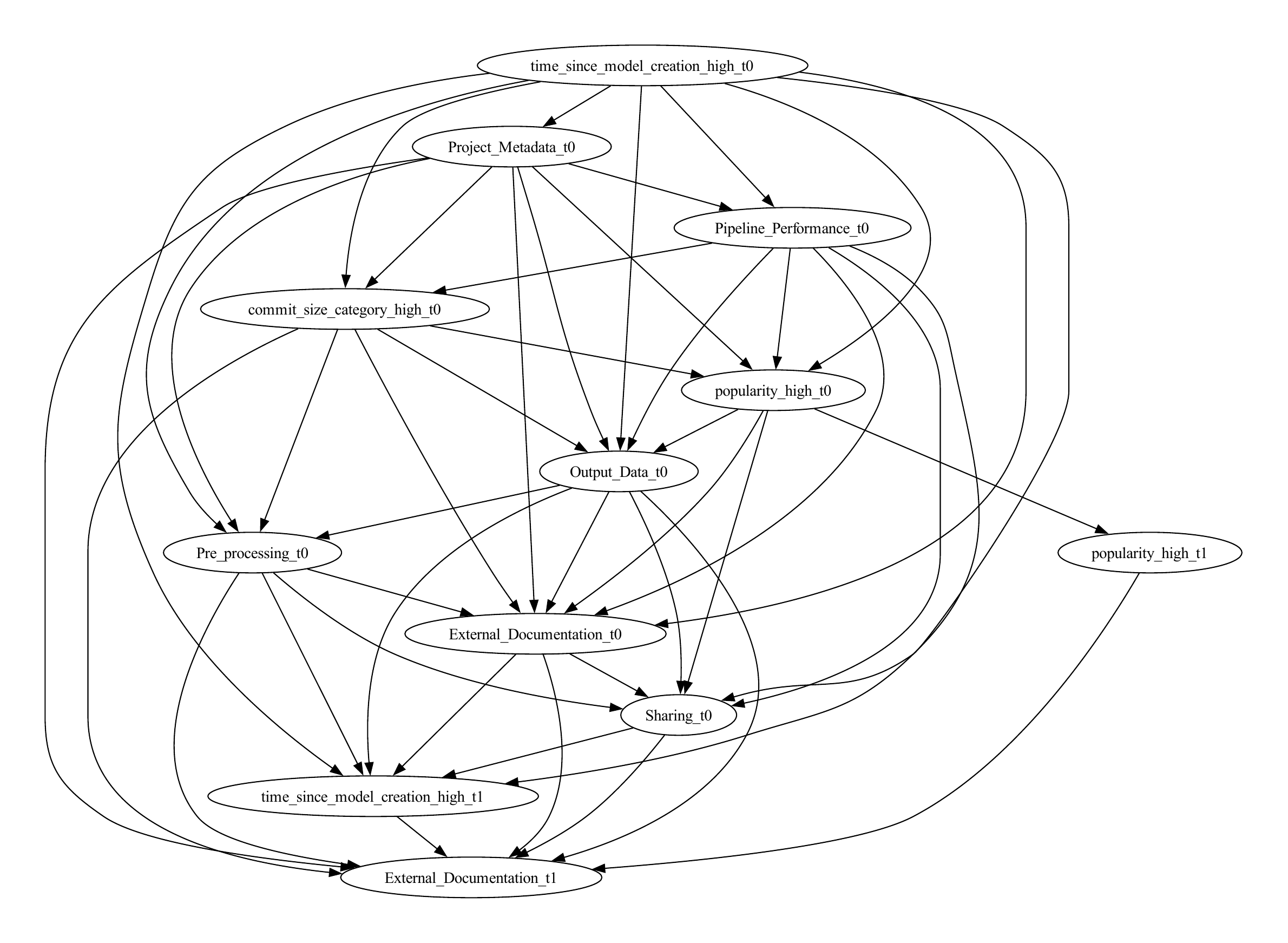}
\caption{Structure of the trained DBN used for to estimate the probability of a commit being of type \textit{External Documentation}}
\label{fig:dbn_ExternalDocumentation1}
\end{figure}

The trained DBN, unrolled for 2 time steps, contains a total of 40 nodes. However, it can be simplified when predicting a specific commit type at the next time step. For example, the DBN used to estimate the probability of a commit being of type \textit{External Documentation}, given all variables from the previous time step, is shown in  Fig.~\ref{fig:dbn_ExternalDocumentation1}.

\subsubsection{Influence of Project Characteristics on Commit Type Dependencies Over Time (RQ2.2)}

\paragraph{\textbf{Relationship between Time Between Commits and Commit Type Dependencies:}}

\begin{figure}[h]
\centering
\includegraphics[width=0.6\textwidth]{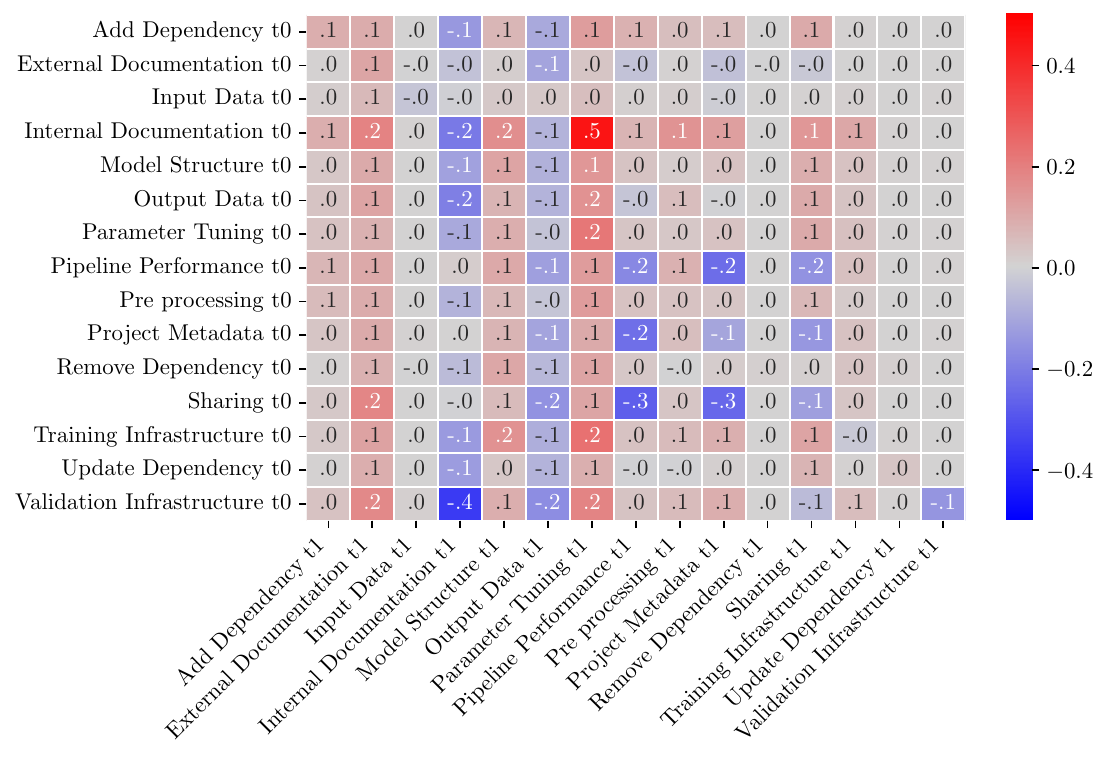}
\caption{Variation in transition probabilities between commits close in time}
\label{fig:RQ2.2_time_between_commits}
\Description{Variation in transition probabilities between commits close in time.}
\end{figure}

Fig. \ref{fig:RQ2.2_time_between_commits} shows the variation in the transition probability between commits that are close in time (less than one day) and those that are not. The time interval between commits significantly influences the type of subsequent activity. Commits made after a long interval are much more likely to involve certain types of focused refinement. For example, if an \textit{Internal Documentation} commit is made, the probability of the next commit being \textit{Parameter Tuning} is dramatically higher when the time interval is long (difference of +0.452). Transitions to \textit{External Documentation} are also consistently more probable after longer intervals. Conversely, rapid, consecutive activities are different. When the time interval is short, transitions related to \textit{Sharing} and \textit{Pipeline Performance} are more common, particularly following metadata or previous sharing commits (e.g., difference of -0.276 for Sharing$_{t0} \rightarrow$ Pipeline\_Performance$_{t1}$; difference of -0.254 for Sharing$_{t0} \rightarrow$ Project\_Metadata$_{t1}$). This suggests a dichotomy between deliberate, spaced-out refinement (tuning, documentation) and rapid-fire iteration on sharing and performance aspects.

\paragraph{\textbf{Relationship between Commit Size and Commit Type Sequences: }}

\begin{figure}[h]
\centering
\includegraphics[width=0.6\textwidth]{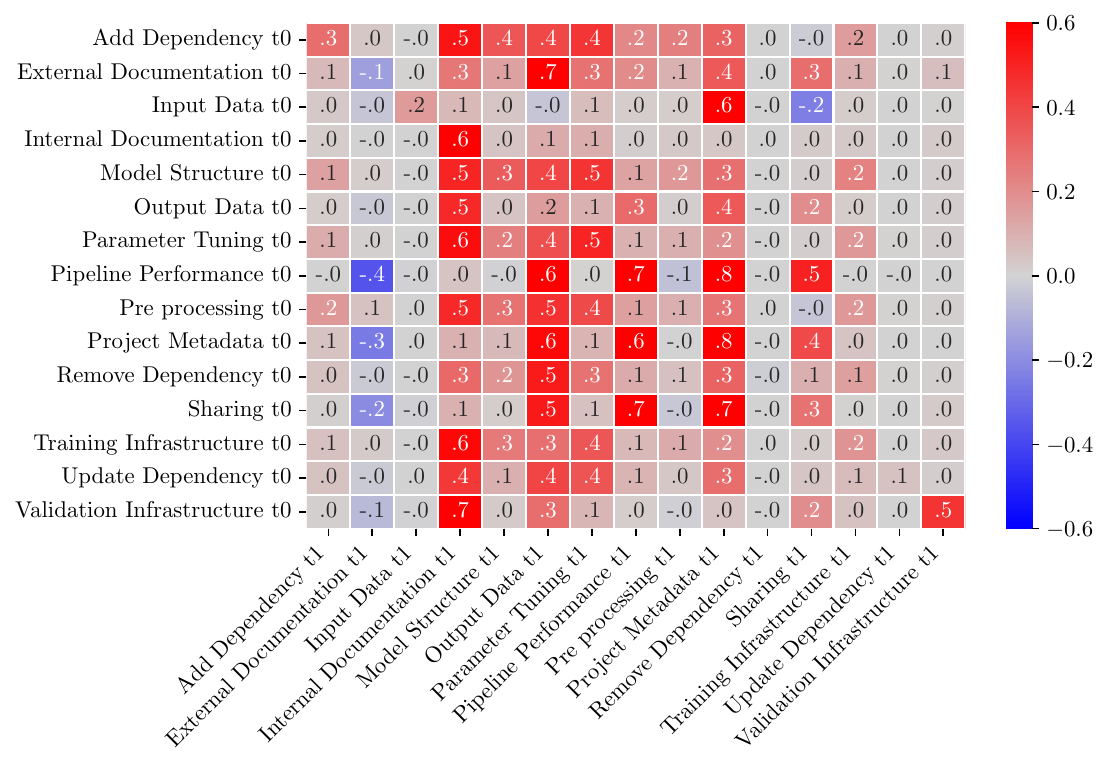}
\caption{Variation in transition probabilities between large and small commits}
\label{fig:RQ2.2_commit_size}
\Description{Variation in transition probabilities between large and small commits.}
\end{figure}

Fig. \ref{fig:RQ2.2_commit_size} shows the variation in transition probability between commits of large and small sizes. Commit size is a strong indicator of the nature of the sequential change. Large commits are significantly more likely to be part of sequences involving major configuration and performance updates. The probability of a \textit{Pipeline Performance} commit being followed by a \textit{Project Metadata} commit is vastly higher when the commits are large (difference of +0.834), as is the probability of a \textit{Sharing} commit being followed by a \textit{Project Metadata} commit (+0.719). Transitions to \textit{Output Data} and \textit{Internal Documentation} are also broadly more probable when the commit size is large. Consecutive \textit{Validation Infrastructure} commits are also much more likely to be large (+0.451). Conversely, transitions to \textit{External Documentation} are often less likely in sequences of large commits (e.g., difference of -0.360 following a \textit{Pipeline Performance} commit), suggesting that large, technically-focused commits are distinct from documentation-focused ones.

\paragraph{\textbf{Relationship between Collaboration Intensity and Commit Type Dependencies: }}

\begin{figure}[h]
\centering
\includegraphics[width=0.6\textwidth]{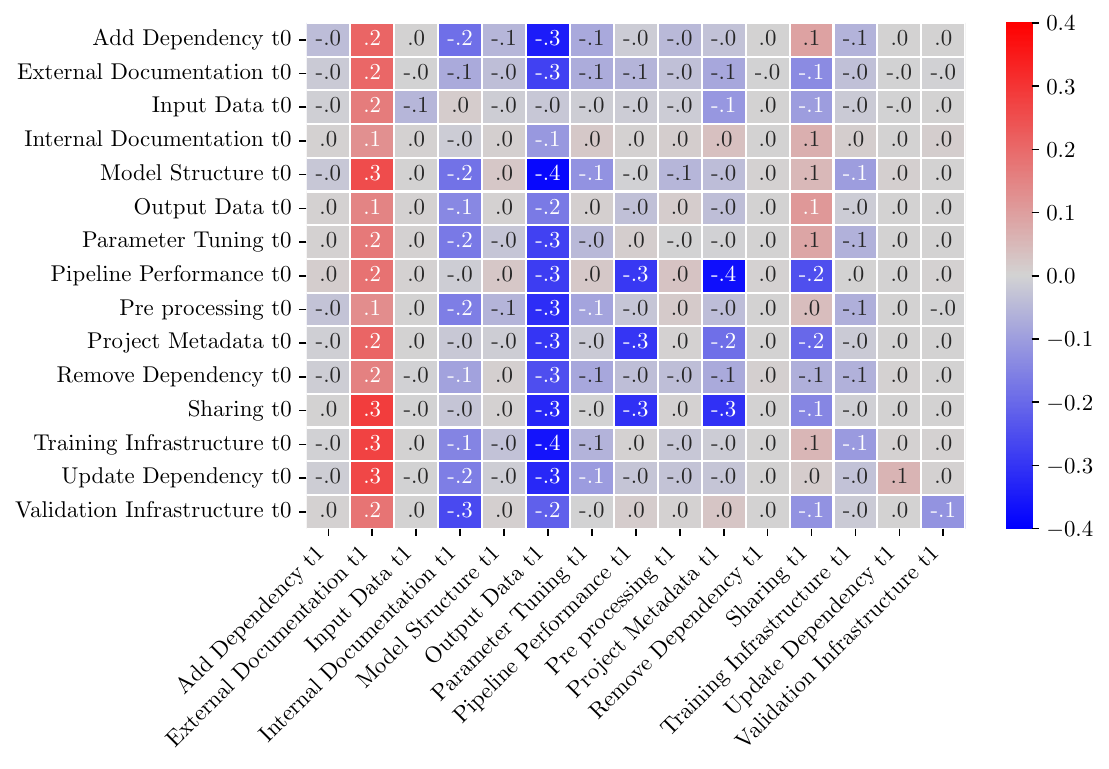}
\caption{Variation in transition probabilities in projects with high and low collaboration intensity}
\label{fig:RQ2.2_collaboration}
\Description{Variation in transition probabilities in projects with high and low collaboration intensity.}
\end{figure}

Fig. \ref{fig:RQ2.2_collaboration} shows the variation in transition probability between projects with high and low collaboration intensity. Collaboration intensity reveals a critical trade-off in development focus. In projects with high collaboration intensity, commit sequences are significantly more likely to involve transitions to \textit{External Documentation} (nearly all differences in that column are positive, e.g., +0.282 following a \textit{Sharing} commit). This indicates a heightened need for communication and coordination through documentation in collaborative settings. Conversely, and strikingly, transitions to \textit{Output Data} are consistently and significantly less likely in highly collaborative projects (nearly all differences in that column are negative, e.g., -0.384 following a \textit{Model Structure} commit; -0.345 following an \textit{Add Dependency} commit). This suggests that in highly collaborative projects, the commit-to-commit evolution prioritizes shared understanding (documentation) over the raw generation of individual artifacts.

\paragraph{\textbf{Relationship between Model Popularity and Commit Type Sequences: }}

\begin{figure}[h]
\centering
\includegraphics[width=0.6\textwidth]{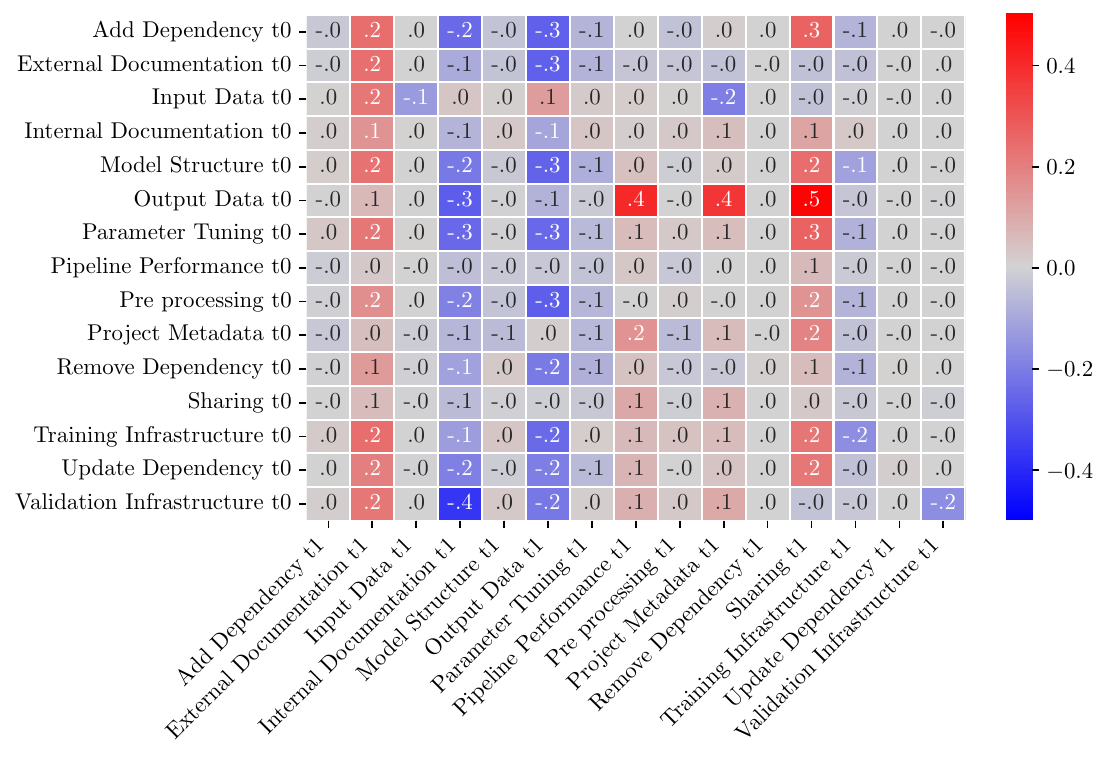}
\caption{Variation in transition probabilities between in popular and non popular projects}
\label{fig:RQ2.2_popularity}
\Description{Variation in transition probabilities between in popular and non popular projects.}
\end{figure}

Fig. \ref{fig:RQ2.2_popularity} shows the variation in transition probability between commits in popular and non-popular projects. Popularity is associated with a distinct evolutionary trajectory. In popular projects, commit sequences are far more likely to feature transitions to \textit{Sharing} (e.g., difference of +0.491 following an \textit{Output Data} commit) and \textit{Pipeline Performance} (+0.399 following an \textit{Output Data} commit). Transitions to \textit{External Documentation} are also generally more probable. In stark contrast, transitions to \textit{Output Data} are consistently less likely in popular projects (most differences in that column are negative, e.g., -0.273 following an \textit{External Documentation} commit). This suggests that the evolution of popular projects is characterized by a focus on dissemination, efficiency, and communication, moving beyond the simple generation of model artifacts that characterizes the sequences in less popular projects.

\begin{myquote}
{\large \textbf{Main Findings for RQ2}}

\vspace{0.2cm}

\textbf{Finding 9}. \textit{Commit sequences demonstrate a temporal clustering of related development activities, evidenced by strong self-dependencies (e.g., consecutive \textit{Output Data}, \textit{Sharing}, or \textit{Pipeline Performance} commits) and logical cross-type transitions (e.g., most development activities are immediately followed by an \textit{Output Data} commit, reflecting a cycle of work followed by artifact generation).}

\vspace{0.2cm}

\textbf{Finding 10}. \textit{Development activities have different temporal pacing. Rapid sequences with short time intervals are more likely to involve \textit{Sharing} and \textit{Pipeline Performance} changes. In contrast, transitions involving \textit{Parameter Tuning} or leading to \textit{External Documentation} are more probable when the time between commits is longer.}

\vspace{0.2cm}

\textbf{Finding 11}. \textit{High collaboration intensity is associated with a significantly increased likelihood of transitions to \textit{External Documentation} commits and a decreased likelihood of transitions to \textit{Output Data} commits, highlighting a shift in evolutionary focus from artifact generation to shared understanding in collaborative settings.} 

\vspace{0.2cm}

\textbf{Finding 12}. \textit{Popular projects evolve distinctly: their commit sequences are significantly more likely to involve transitions to \textit{Sharing}, \textit{Pipeline Performance}, and \textit{External Documentation}. Conversely, transitions that simply generate new \textit{Output Data} are less common, suggesting their evolution prioritizes efficiency and dissemination over raw artifact production.}
\end{myquote}

\subsection{RQ3: Analysis of Release Types and Patterns in models}

\subsubsection{Distribution and Evolution of Release Types (RQ3.1)}

To understand the landscape of releases in models, we analyzed the distribution and evolution of release types across the HF ecosystem using data from 202 models with over 2,251 tagged releases in total.
\paragraph{\textbf{Distribution of Release Types:}}

The analysis reveals that \textit{Output Data} (1,544 releases), \textit{Sharing} (1,446 releases), and \textit{External Documentation} (1,355 releases) are the three most frequent types associated with releases. This suggests that releases are primarily used to publish new model artifacts, facilitate their dissemination, and communicate changes to end-users. This contrasts with the commit-level data, where Project Metadata was more prominent than External Documentation. Releases involving \textit{Project Metadata} (494), \textit{Training Infrastructure} (458), \textit{Parameter Tuning} (457), and \textit{Internal Documentation} (449) are also notably present.

Conversely, the least common release types are \textit{Remove Dependency} (0 releases), \textit{Add Dependency} (3 releases), and \textit{Input Data} (31 releases). The extreme rarity of dependency changes in tagged releases reinforces the idea that dependency structures are highly stable or that such changes are not considered release-worthy milestones on their own.

\paragraph{\textbf{Evolution of Release Types Over Time:}}

Fig.~\ref{fig:RQ3.1_Evolution_Over_Time} illustrates the evolution of release types over time up to Q1 2025. The graph reveals several trends:

\begin{figure}[h]
\centering
\includegraphics[width=0.6\textwidth]{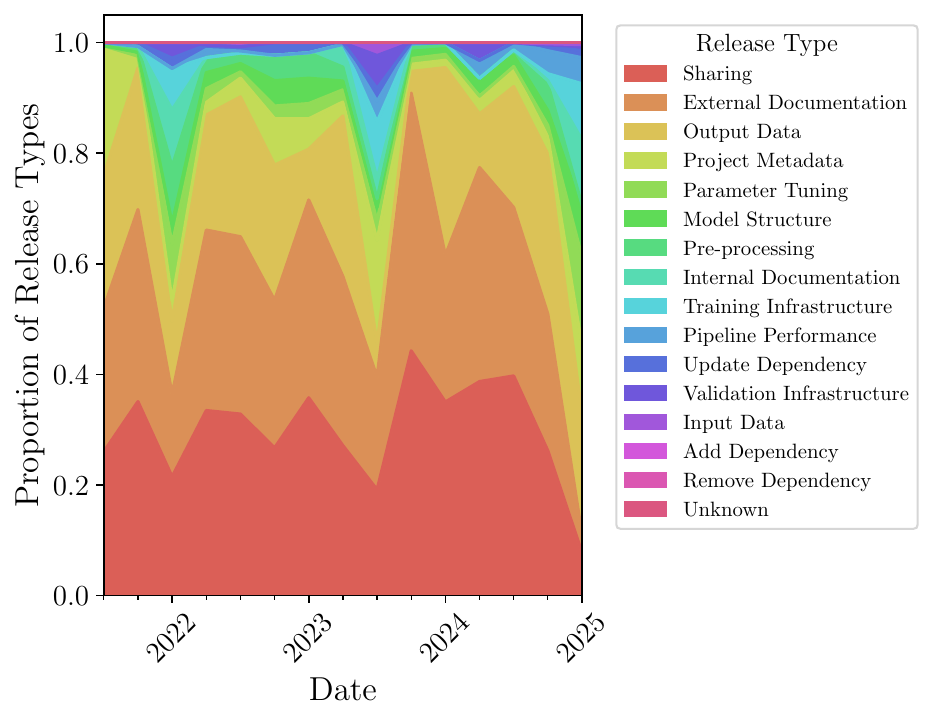} 
\caption{Evolution of Release Types Over Time}
\label{fig:RQ3.1_Evolution_Over_Time}
\Description{Evolution of Release Types Over Time}
\end{figure}

\begin{itemize}
    \item \textit{Sharing} and \textit{External Documentation} have been the most dominant release types throughout most of the platform's history, often comprising over 30-40\% of releases in any given quarter (e.g., Sharing at 0.443 and External Documentation at 0.466 in 2023Q4). This underscores the central role of releases as communication and dissemination tools.
    \item \textit{Output Data} has been consistently prevalent, often being the third most common release type, highlighting that publishing new model weights is a core function of a release.
    \item The most recent period (Q1 2025) shows a notable shift, with a sharp increase in the proportion of releases related to \textit{Parameter Tuning} (0.161), \textit{Internal Documentation} (0.118), and \textit{Training Infrastructure} (0.101), while the proportion of \textit{Sharing} and \textit{External Documentation} decreased. This could signal a recent trend toward more technically focused, iterative releases.
    \item In contrast to its prevalence in commits, \textit{Project Metadata} constitutes a smaller, though consistent, proportion of release types.
\end{itemize}

The historical dominance of \textit{Sharing} and \textit{External Documentation} in releases confirms their function as major communication milestones. The recent uptick in more technical release types like \textit{Parameter Tuning} and \textit{Training Infrastructure} may reflect evolving practices towards more frequent, granular versioning of the entire ML pipeline.

\paragraph{\textbf{Release Types Across Project Phases:}}

\begin{figure}[h]
\centering
\includegraphics[width=0.75\textwidth]{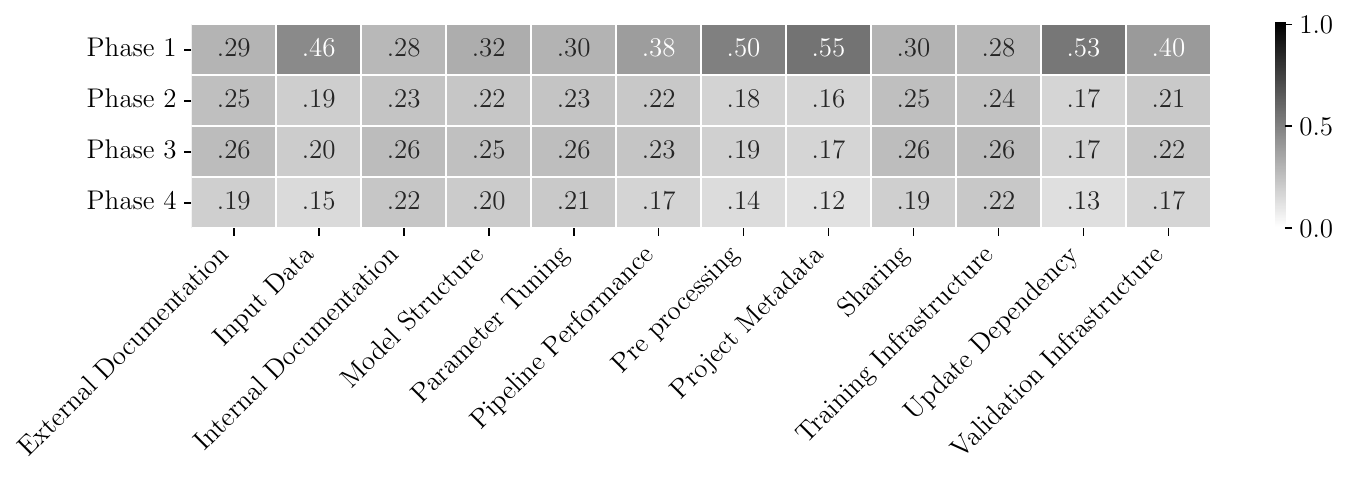} 
\caption{Distribution of Release Types Across Project Phases}
\label{fig:RQ3.1_Phases}
\Description{Distribution of Release Types Across Project Phases}
\end{figure}

Fig. \ref{fig:RQ3.1_Phases} shows the distribution of release types across different project phases (quartiles of a model's release history). The initial phase (quartile 1) of a project's release history concentrates a wide variety of foundational activities. The highest probabilities in this phase are for \textit{Project Metadata} (0.550), \textit{Update Dependency} (0.532), and \textit{Pre-processing} (0.500). Following closely are \textit{Input Data} (0.459) and \textit{Validation Infrastructure} (0.397). This indicates that early releases are often concerned with setting up the repository, defining dependencies, and establishing the data and validation pipelines. Most other types, including \textit{Model Structure} (0.323) and \textit{Parameter Tuning} (0.297), also show their highest probability in Phase 1, though less pronounced. The general trend is for the probability of nearly all release types to decrease after the first phase, underscoring that initial releases are often comprehensive, bundling many different kinds of changes, while later releases may be more focused. (Note: \textit{Add/Remove Dependency} are not prominently featured due to having almost no occurrences in the release dataset).

\subsubsection{Association of Project Characteristics with Release Types (RQ3.2)}

Our analysis of how project characteristics influence release types throughout a model's lifecycle revealed several interesting patterns.
\paragraph{\textbf{Correlations between Project Characteristics and Release Types:}}

\begin{figure}[h]
\centering
\includegraphics[width=0.6\textwidth]{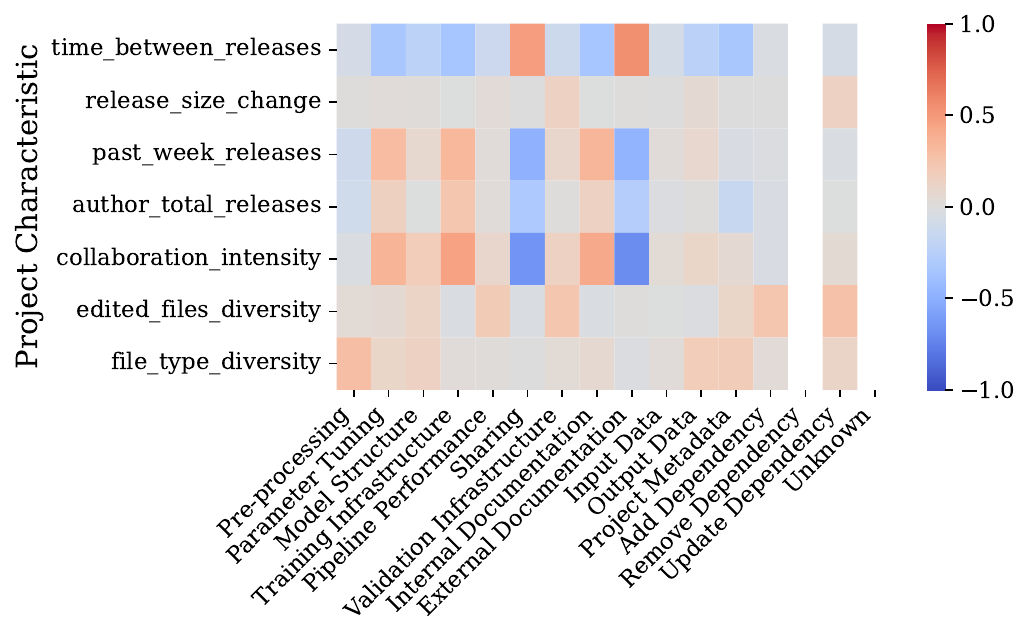} 
\caption{Correlations between Project Characteristics and Release Types}
\label{fig:RQ3.2_Correlations}
\Description{Correlations between Project Characteristics and Release Types}
\end{figure}

Fig. \ref{fig:RQ3.2_Correlations} presents the correlations between project characteristics and release types. The correlations reveal a clear dichotomy in release strategies. On one hand, long intervals between releases (\textit{time\_between\_releases}) are strongly and positively correlated with releases focused on communication and dissemination: \textit{External Documentation} (+0.550) and \textit{Sharing} (+0.484). On the other hand, high-frequency, collaborative development is linked to more technical releases. \textit{Collaboration intensity} has strong positive correlations with \textit{Training Infrastructure} (+0.459), \textit{Internal Documentation} (+0.417), and \textit{Parameter Tuning} (+0.352). Correspondingly, \textit{collaboration\_intensity} is strongly and negatively correlated with \textit{External Documentation} (-0.691) and \textit{Sharing} (-0.649). This suggests two distinct modes of operation: one involving infrequent, major releases focused on public communication, and another characterized by rapid, collaborative cycles that produce more technically-focused releases (e.g., updating training logic or parameters).

\paragraph{\textbf{Release Types Across Model Sizes:}}

\begin{figure}[h]
\centering
\includegraphics[width=0.6\textwidth]{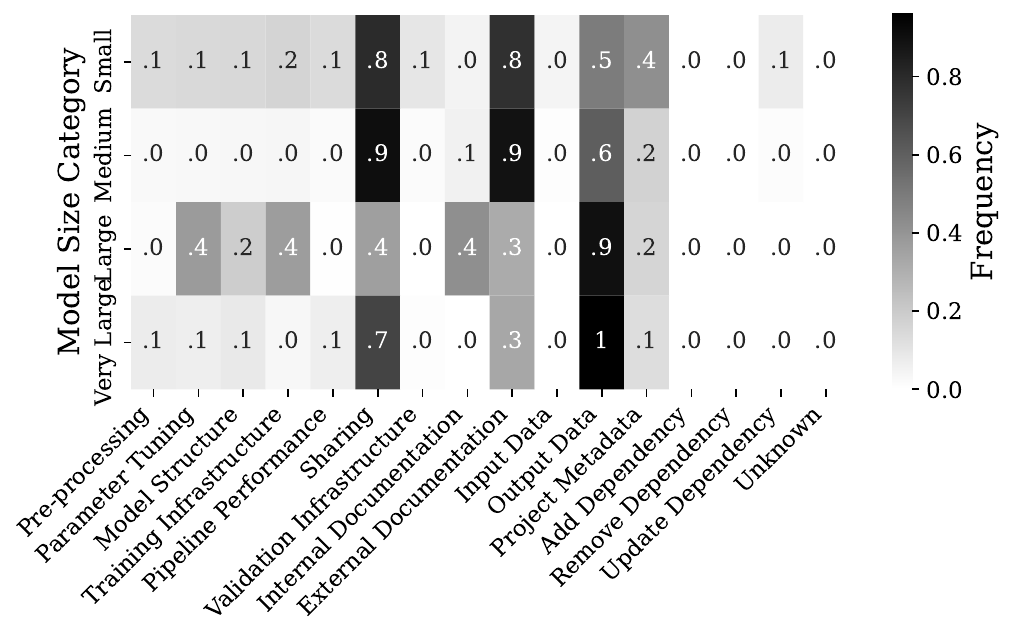} 
\caption{Release Type Distribution by Model Size}
\label{fig:RQ3.2_Model_Size_Distribution}
\Description{Release Type Distribution by Model Size}
\end{figure}

The distribution of release types across different model sizes, as shown in Fig. \ref{fig:RQ3.2_Model_Size_Distribution}, reveals a clear shift in focus as models scale. Releases for Small and Medium models are dominated by communication and dissemination. For Medium models, the mean frequency of \textit{Sharing} is 0.903 and \textit{External Documentation} is 0.888. This indicates that for smaller projects, a release is primarily an act of communication.
In stark contrast, for Large and Very Large models, the most frequent release type by a wide margin is \textit{Output Data} (mean frequencies of 0.894 and 0.957, respectively). This suggests that as models become larger and potentially more central to a project, the act of releasing becomes primarily about publishing the new, tangible model artifact itself. While communication is still present, the focus of the release event shifts to the data product.

\paragraph{\textbf{Release Types by Time Between Releases:}}

\begin{figure}[h]
\centering
\includegraphics[width=0.6\textwidth]{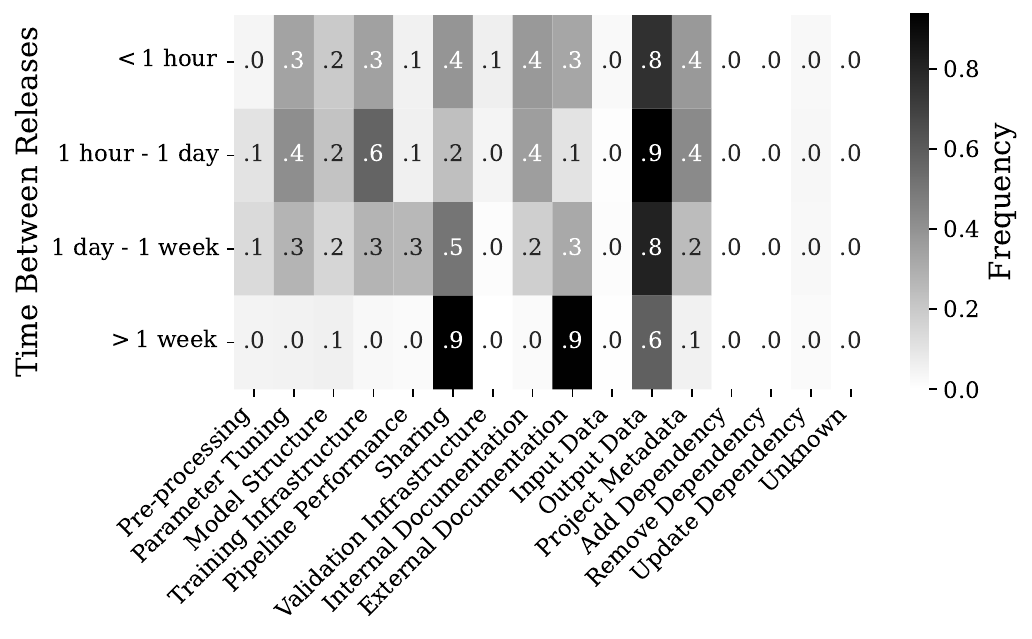} 
\caption{Release Type Distribution by Time Between Releases}
\label{fig:RQ3.2_Time_Between_Releases}
\Description{Release Type Distribution by Time Between Releases}
\end{figure}

The relationship between release types and the time interval between releases, shown in Fig. \ref{fig:RQ3.2_Time_Between_Releases}, strongly reinforces the dichotomy between rapid, technical updates and deliberate, communicative milestones. Releases made after a long interval (more than a week) are overwhelmingly characterized by \textit{Sharing} (mean frequency of 0.933) and \textit{External Documentation} (0.931). This confirms that major, infrequent releases are planned communication events.
Conversely, rapid release cadences are associated with technical changes. For releases made very shortly after the previous one (<1 hour), the most frequent types are \textit{Output Data} (0.759), \textit{Project Metadata} (0.374), \textit{Internal Documentation} (0.376), \textit{Training Infrastructure} (0.346), and \textit{Parameter Tuning} (0.339). This indicates that frequent, iterative releases often bundle the products of development cycles (new weights, logs, and parameters) and associated configuration updates.

\subsubsection{Patterns in the Evolution of Release Types (RQ3.3)}

\paragraph{\textbf{Dependencies Between Consecutive Releases:}}

\begin{figure}[h]
\centering
\includegraphics[width=0.65\textwidth]{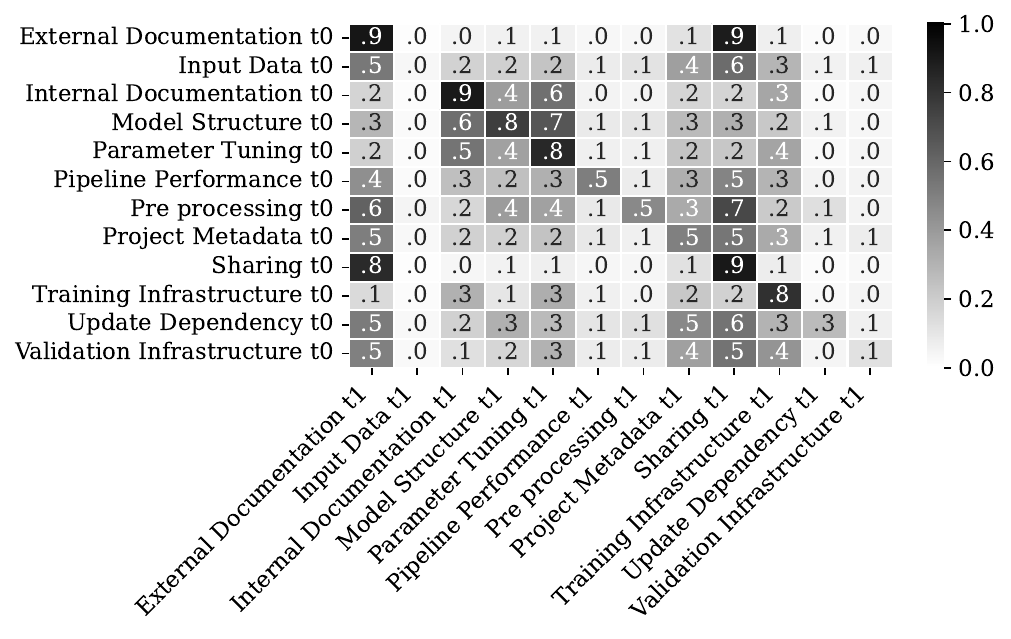} 
\caption{Probability of Consecutive Releases Types}
\label{fig:RQ3.3_Consecutive_Releases}
\Description{Probability of Consecutive Releases Types}
\end{figure}

Fig. \ref{fig:RQ3.3_Consecutive_Releases} illustrates the probability of each release type being followed by another based on the DBN analysis. The data reveals a strong tendency for development focus to persist across consecutive releases. Extremely high self-transition probabilities are observed for key release types, notably \textit{External Documentation} (0.913), \textit{Sharing} (0.904), \textit{Internal Documentation} (0.900), \textit{Parameter Tuning} (0.843), and \textit{Training Infrastructure} (0.812). This indicates that when a release focuses on one of these aspects, the immediate next release is highly likely to continue that theme, suggesting iterative refinement or phased rollouts.

Beyond self-transitions, several cross-type transition probabilities highlight logical development workflows. The strongest such link is from \textit{Sharing} to \textit{External Documentation} (0.888), confirming that dissemination efforts are almost always followed by documentation updates. Architectural changes (\textit{Model Structure}) are frequently followed by releases focused on \textit{Parameter Tuning} (0.662) and \textit{Internal Documentation} (0.569), representing a logical flow from changing the architecture to tuning it and logging the results. Similarly, a \textit{Parameter Tuning} release is often followed by an \textit{Internal Documentation} release (0.546). These sequences underscore a development process where core technical changes are followed by periods of refinement and artifact generation.

\paragraph{\textbf{Influence of Time Between Releases:}}

\begin{figure}[h]
\centering
\includegraphics[width=0.65\textwidth]{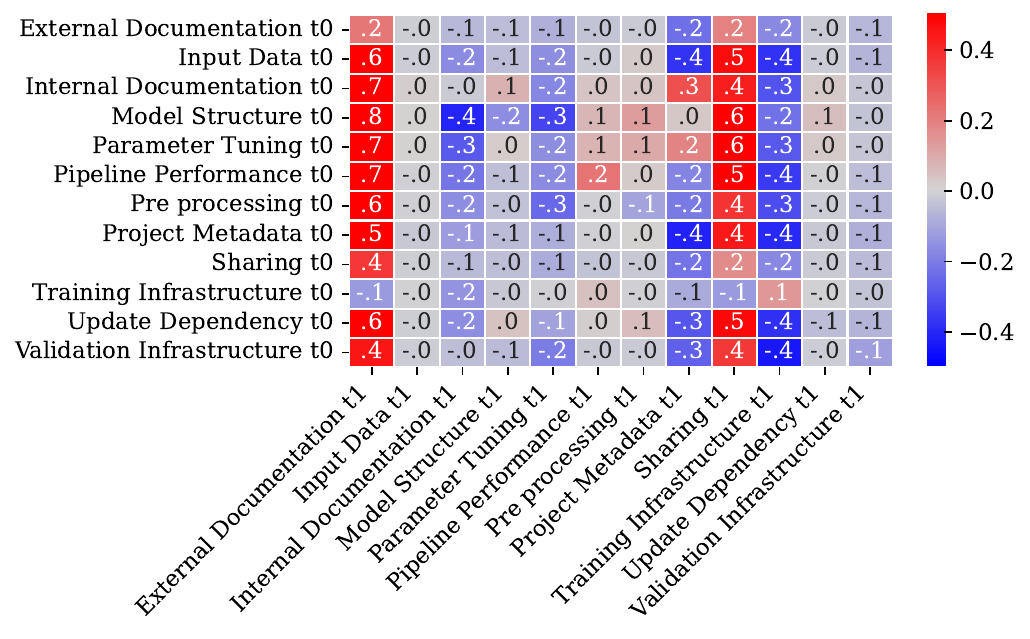} 
\caption{Variation in Transition Probabilities by Time Between Releases}
\label{fig:RQ3.3_Time_Between_Releases}
\Description{Variation in Transition Probabilities by Time Between Releases}
\end{figure}

Fig. \ref{fig:RQ3.3_Time_Between_Releases} shows the variation in transition probabilities between releases made close in time and those further apart. The data reveals a clear separation between rapid, iterative releases and major, milestone releases. When the time interval between releases is long, transitions to communicative types like \textit{External Documentation} and \textit{Sharing} are far more probable (e.g., the difference in probability for a \textit{Model Structure} release to be followed by an \textit{External Documentation} release is +0.752 for long intervals). Conversely, when the time interval is short, release sequences are dominated by technical updates. Transitions to \textit{Training Infrastructure}, \textit{Parameter Tuning}, and \textit{Project Metadata} are all significantly more likely in rapid release cycles (indicated by large negative differences, e.g., -0.442 for VI$_{t0} \rightarrow$ TI$_{t1}$; -0.420 for PJ$_{t0} \rightarrow$ PJ$_{t1}$). This suggests long periods facilitate major documentation and sharing efforts, while shorter cycles are characterized by the rapid iteration of technical components.

\paragraph{\textbf{Influence of Release Size:}}

\begin{figure}[h]
\centering
\includegraphics[width=0.6\textwidth]{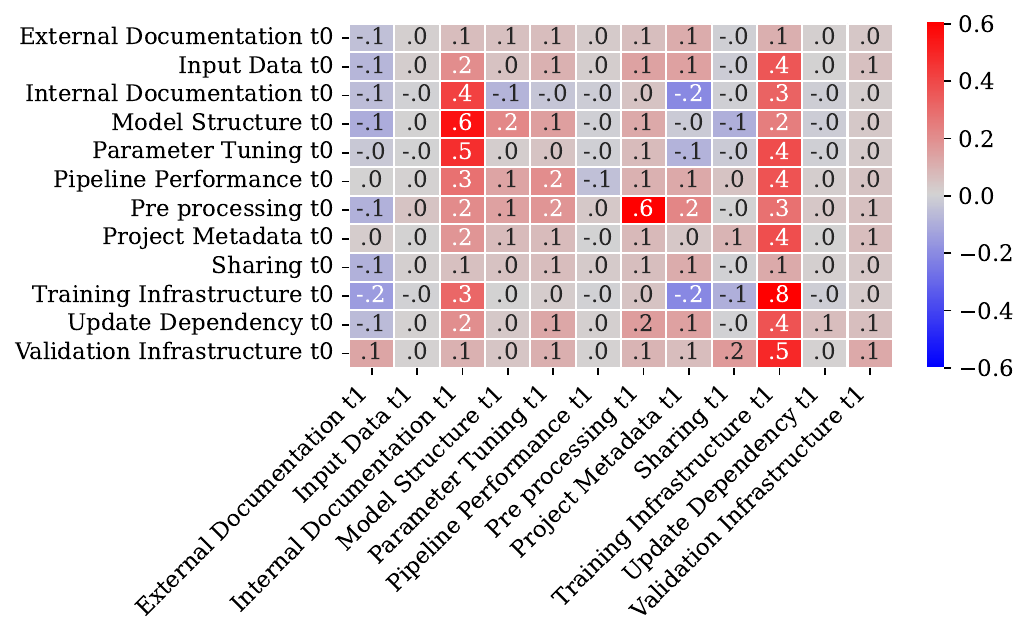} 
\caption{Variation in Transition Probabilities by Release Size}
\label{fig:RQ3.3_Release_Size}
\Description{Variation in Transition Probabilities by Release Size}
\end{figure}

Fig. \ref{fig:RQ3.3_Release_Size} shows the variation in transition probabilities between large and small releases. Large releases are strongly associated with sequences of intensive technical work. The probability of two consecutive \textit{Training Infrastructure} releases is dramatically higher when the releases are large (difference of +0.826). More broadly, transitions to \textit{Training Infrastructure} and \textit{Internal Documentation} are consistently more likely when the release size is high, regardless of the preceding release type. For instance, a \textit{Parameter Tuning} release is much more likely to be followed by a \textit{Training Infrastructure} (+0.381) or \textit{Internal Documentation} (+0.466) release if the releases are large. This implies that large release packages are indicative of significant advancements in the training systems and the comprehensive bundling of internal artifacts, rather than just documentation updates.

\paragraph{\textbf{Influence of Collaboration Intensity:}}

\begin{figure}[h]
\centering
\includegraphics[width=0.6\textwidth]{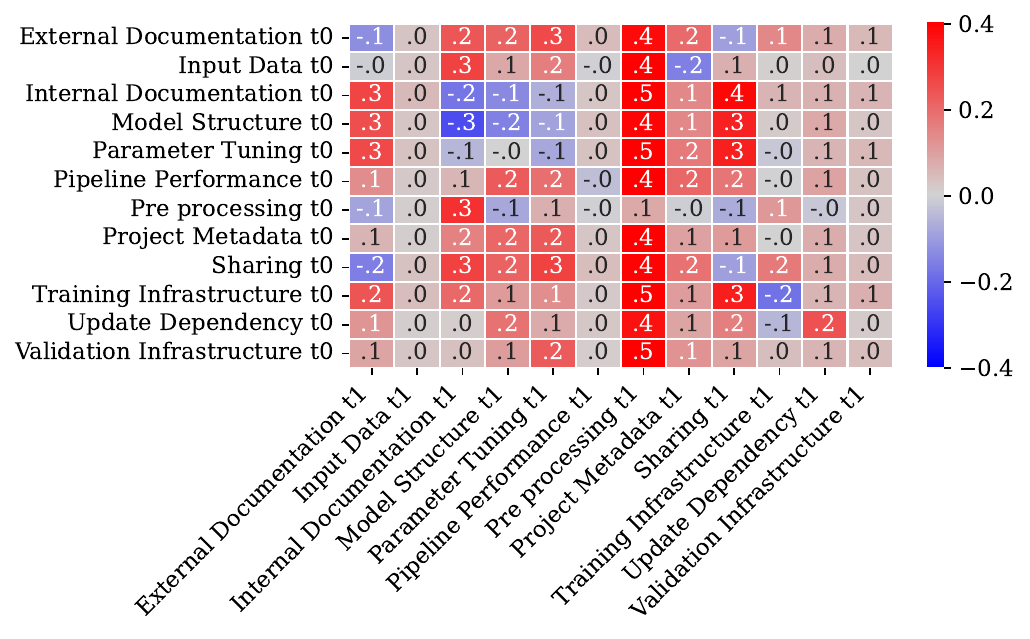} 
\caption{Variation in Transition Probabilities by Collaboration Intensity}
\label{fig:RQ3.3_Collaboration_Intensity}
\Description{Variation in Transition Probabilities by Collaboration Intensity}
\end{figure}

Fig. \ref{fig:RQ3.3_Collaboration_Intensity} shows the variation in transition probabilities in projects with high and low collaboration intensity. Highly collaborative projects exhibit more technically dense release sequences. In projects with high collaboration intensity, there is an increased likelihood for transitions to technical release types such as \textit{Training Infrastructure}, \textit{Parameter Tuning}, \textit{Model Structure}, and \textit{Pre-processing} (most differences in these columns are positive). For example, a \textit{Sharing} release is more likely to be followed by a \textit{Training Infrastructure} release (+0.169) in a highly collaborative environment. These patterns suggest that highly collaborative environments prioritize releases that reflect the active co-development and evolution of core model components and the underlying training processes.

\paragraph{\textbf{Influence of Project Popularity:}}

\begin{figure}[h]
\centering
\includegraphics[width=0.6\textwidth]{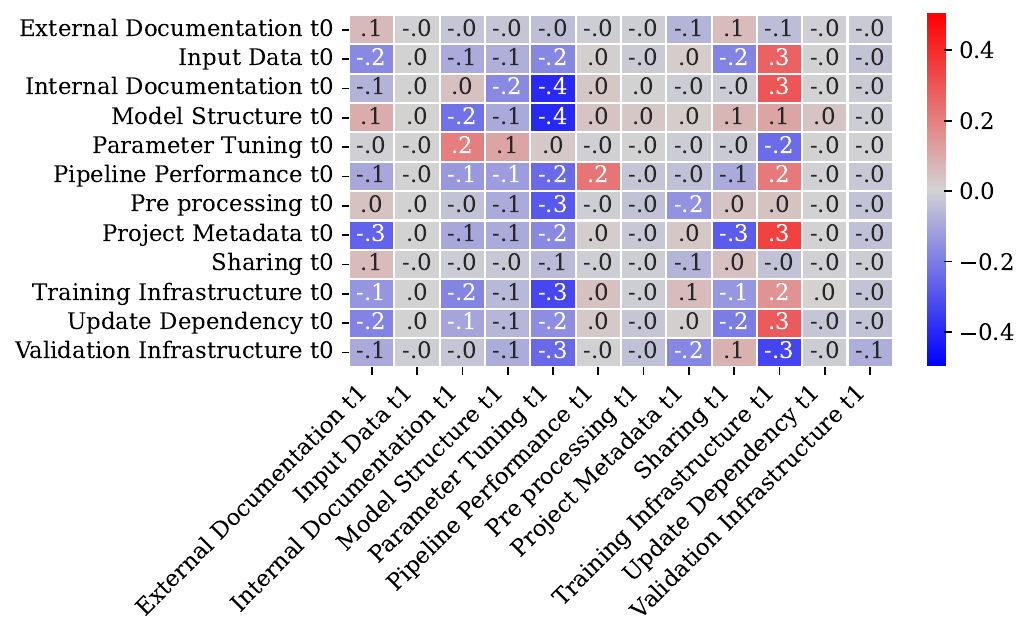} 
\caption{Variation in Transition Probabilities by Project Popularity}
\label{fig:RQ3.3_Popularity}
\Description{Variation in Transition Probabilities by Project Popularity}
\end{figure}

Fig. \ref{fig:RQ3.3_Popularity} shows the variation in transition probabilities in popular and non-popular projects. Popularity is associated with a distinct shift in release evolution. Popular projects are significantly more likely to feature release sequences involving \textit{Training Infrastructure} updates (e.g., a \textit{Project Metadata} release is far more likely to be followed by a \textit{Training Infrastructure} release, with a difference of +0.347). Conversely, transitions to \textit{Parameter Tuning} and \textit{External Documentation} are generally less likely in the release sequences of popular projects (e.g., a \textit{Model Structure} release is much less likely to be followed by a \textit{Parameter Tuning} release, with a difference of -0.403). This suggests that as projects gain popularity, their release strategy may mature from frequent tuning and documentation cycles towards more substantial, infrastructure-focused updates, possibly reflecting a move towards greater stability and scalability.

\subsubsection{Analysis of Metadata Changes (RQ3.4)}

Our analysis of metadata changes in 28 models reveals that between releases, ML repositories primarily adjust numerical values of weights rather than changing architecture, adding new parameters, or altering tensor shapes. This indicates that most changes focus on weight adjustments rather than structural modifications.

\begin{myquote}
{\large \textbf{Main Findings for RQ3}}

\vspace{0.2cm}

\textbf{Finding 13}. \textit{Releases tend to consolidate significant updates, particularly in \textit{Output Data}, \textit{Sharing}, and \textit{External Documentation}, serving as key development milestones. This distinguishes their role from the more granular, frequent changes (dominated by \textit{Output Data}, \textit{Project Metadata}, and \textit{Sharing}) observed in individual commits.}

\vspace{0.2cm}

\textbf{Finding 14}. \textit{Release patterns vary with collaboration intensity. Highly collaborative projects feature more technically dense release sequences, with more frequent transitions to \textit{Training Infrastructure}, \textit{Parameter Tuning}, and \textit{Model Structure} releases, reflecting a focus on co-development of core components.}

\vspace{0.2cm}

\textbf{Finding 15}. \textit{Release sequences show that key updates (\textit{External Documentation, Sharing, Internal Documentation, Parameter Tuning, Training Infrastructure}) often persist across consecutive releases. Common cross-type transitions, like \textit{Sharing} leading to \textit{External Documentation} and \textit{Model Structure} changes leading to \textit{Parameter Tuning}, reveal logical development and consolidation phases.}

\vspace{0.2cm}

\textbf{Finding 16}. \textit{Metadata changes between releases primarily involve numerical weight adjustments rather than architectural modifications, indicating a focus on fine-tuning over structural changes.}
\end{myquote}

\section{DISCUSSIONS AND IMPLICATIONS} \label{sec:discussions}

Our large-scale, longitudinal analysis of model changes on the HF platform reveals several significant patterns, offering insights into how ML models evolve differently from traditional software and what this implies for development practices. This section discusses these findings by contrasting ML model evolution with traditional software, interpreting patterns related to data science methodologies, model popularity, and collaboration, and proposing evidence-grounded implications.

\subsection{ML Model Evolution vs. Traditional Software Evolution}

Understanding the evolution of ML models requires recognizing its distinctions from traditional software evolution. While both involve code, documentation, and infrastructure, the nature and emphasis of changes differ significantly.

\textbf{Nature of Artifacts and Changes:} Traditional software evolution primarily revolves around changes to source code logic, user interfaces, and feature sets \cite{swanson1976dimensions, hindle2008large}. In contrast, ML model evolution involves a broader set of artifacts. While code is important, our findings show that the most frequent changes revolve around data and configuration.
\begin{itemize}
    \item \textit{Data and Configuration as Primary Artifacts:} The most frequent commit types are \textit{Output Data}, \textit{Project Metadata}, and \textit{Sharing} (Finding 1). This is a stark difference from traditional software where code logic changes would dominate. Here, the "executable" (the model weights, a form of \textit{Output Data}) and repository-level configurations (\textit{Project Metadata}) are the most frequently modified artifacts. This focus on data artifacts intensifies with model scale, as releases for large models are overwhelmingly dominated by new \textit{Output Data} (Figure~\ref{fig:RQ3.2_Model_Size_Distribution}). Furthermore, our analysis of metadata changes (Finding 16) confirms that releases often involve numerical weight adjustments rather than just code changes.
    \item \textit{The Role of Infrastructure:} While specialized infrastructure is crucial for ML, our new data shows that direct commits to \textit{Training Infrastructure} are proportionally less frequent than the management of data and metadata artifacts (Finding 1, Finding 2). This suggests that while infrastructure is a key enabler, much of the day-to-day evolution captured in commits is focused on running experiments and managing their outputs, rather than constantly re-architecting the training process itself.
\end{itemize}

\textbf{Development Process and Iteration:} The ML development lifecycle, as our findings suggest an alignment with CRISP-DM (discussed further in Section \ref{ssec:crispdm_alignment}), is inherently more empirical and iterative than many traditional software processes. This is evidenced by:
\begin{itemize}
    \item \textit{Intense, Bundled Experimentation:} The strong co-occurrence of commit types (Finding 8), where adding a dependency is almost always bundled with changes to \textit{Model Structure}, \textit{Parameter Tuning}, and \textit{Output Data},  reveals that development often happens in comprehensive, multi-faceted updates, not small, isolated code changes.
    \item \textit{A Cycle of Work and Artifact Generation:} The DBN analysis (Finding 9) shows that a vast array of activities (tuning, performance work, documentation) are almost immediately followed by an \textit{Output Data} commit. This paints a picture of a development process defined by a tight loop: perform an action, run the pipeline, commit the resulting artifact.
\end{itemize}

\textbf{Release Content and Purpose:} In traditional software, releases often bundle new features. Our findings paint a different primary picture for ML model releases on HF. The prevalence of \textit{Output Data}, \textit{Sharing}, and \textit{External Documentation} as the most frequent release types (Finding 13) indicates that a release is primarily an act of publishing a new artifact and communicating it to the world. This is distinct from a feature-driven release and aligns more with the concept of releasing a new, improved dataset or scientific finding.

\subsection{Alignment with Data Science Methodologies and Lifecycle Phases}
\label{ssec:crispdm_alignment}

Our findings on commit type distributions across project phases (Finding 3) resonate strongly with established data science methodologies like CRISP-DM \cite{Chapman2000CRISPDM1S} and the iterative nature described by Data Science Trajectories (DST) \cite{DSTCrisp}. This alignment itself is a key characteristic that distinguishes ML development from more linear or feature-gated traditional software lifecycles.

\begin{figure}[h]
\centering
\includegraphics[width=\textwidth]{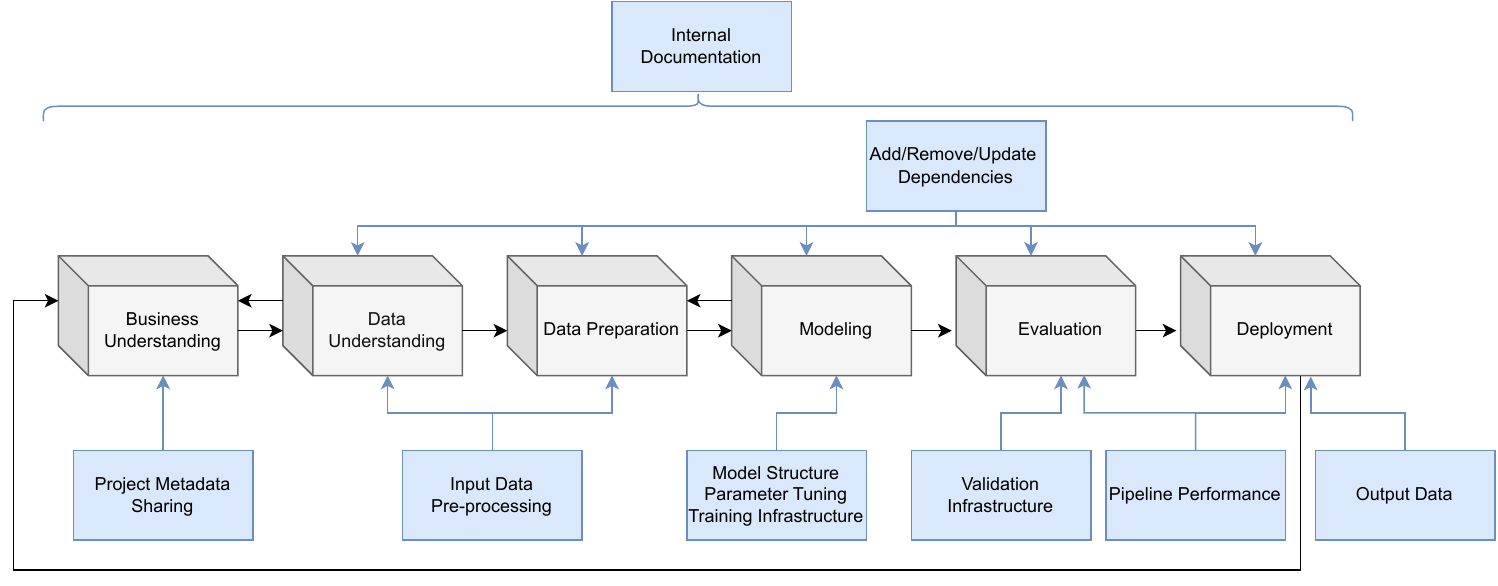}
\caption{Mapping of Commit Types to CRISP-DM Phases}
\label{fig:CRISP-DM_diagram}
\Description{Mapping of Commit Types to CRISP-DM Phases}
\end{figure}

\paragraph{Mapping Commit Types to CRISP-DM Phases}
By analyzing the distribution of commit types across project phases (approximated by commit count quartiles, Figure \ref{fig:commit_types}), we observe patterns corresponding to CRISP-DM phases (Figure \ref{fig:CRISP-DM_diagram}).
\begin{itemize}
    \item \textbf{Initial Setup (Business/Data Understanding):} The predominance of \textit{Project Metadata} commits in Phase 1 (Finding 3) aligns with initial project setup, akin to defining objectives and understanding data requirements.
    \item \textbf{Iterative Development (Data Prep, Modeling, Evaluation):} The concentration of diverse activities like \textit{Model Structure}, \textit{Parameter Tuning}, \textit{Pre-processing}, and \textit{Training Infrastructure} commits in Phase 3 (Finding 3) reflects the intense, iterative cycles of data preparation, modeling, and evaluation inherent in ML. This iterative experimentation is a core differentiator from many traditional software development workflows.
    \item \textbf{Consolidation/Deployment Preparation (Deployment):} The overall high frequency of \textit{Output Data} and \textit{Sharing} commits (Finding 1), and the tendency for \textit{External Documentation} updates to occur after longer intervals (Finding 6), suggest a phase focused on finalizing model products and preparing them for wider use, akin to the deployment phase.
\end{itemize}

\paragraph{Implications for ML Development Practices}
Recognizing these patterns offers specific guidance for ML projects:
\begin{itemize}
    \item \textbf{Phase-Specific Resource Allocation:} Understanding that foundational \textit{Project Metadata} work is crucial early, while intensive experimentation (\textit{Model Structure, Parameter Tuning}) often peaks mid-lifecycle (Finding 3), can help teams allocate resources more effectively.
    \item \textbf{Embrace and Tool for Bundled, Iterative Workflows:} The new findings show that ML development proceeds in cohesive, focused bursts. The extreme bundling of tasks in single commits (Finding 8) and the powerful `work -> artifact` cycle seen in commit sequences (Finding 9) demand MLOps tooling and practices that support and track these multi-faceted experiments as single units. Version control must handle not just code, but configurations, parameters, and data artifacts as an atomic experimental unit.
    \item \textbf{Adapt Development Focus to Model Scale:} The clear shift in commit patterns based on model size (Finding 5) provides a roadmap for adapting priorities. For smaller projects, the focus is on the core loop of generating artifacts (\textit{Output Data}, \textit{Internal Documentation}). For very large models, the focus must expand significantly to include \textit{Pipeline Performance}, public-facing management (\textit{Sharing}), and repository configuration (\textit{Project Metadata}). This suggests that scaling an ML project is not just about more data or bigger models, but about a fundamental shift in engineering priorities towards efficiency and governance.
\end{itemize}

\subsection{Model Popularity and Early Development Choices}

Our findings suggest a correlation between early development patterns and eventual model popularity, offering a revised perspective on what might lead to success.
Finding 7 reveals that popular projects, in their initial phase, have a significantly lower proportion of foundational commits like \textit{Project Metadata}, \textit{Model Structure}, and \textit{Parameter Tuning}. This counters the intuition that successful projects start with meticulous, incremental setup. Instead, it suggests that popular models may begin from a more mature or templated starting point, with foundational work completed "off-screen" before the first public commits.
Furthermore, the evolutionary analysis of commit sequences (Finding 12) shows that once established, popular projects evolve differently. Their commit-to-commit changes are significantly more likely to involve \textit{Sharing}, \textit{Pipeline Performance}, and \textit{External Documentation}, and less likely to be simple \textit{Output Data} generation.
Taken together, a new picture emerges: successful projects may start clean and complete, and their ongoing maintenance prioritizes efficiency, communication, and dissemination over the raw experimental churn that characterizes less popular projects. This implies that for a project to gain traction, establishing a solid, well-documented base and then focusing on performance and user engagement may be a more effective strategy than perpetual, uncommunicated experimentation.

\subsection{Collaboration Intensity: Tailoring Practices for Different Team Structures}
\label{ssec:collaboration_implications}

Our study reveals that collaboration intensity is associated with distinct commit and release patterns (Findings 4, 11, 14), prompting a nuanced discussion beyond universally prescribing certain behaviors as "good practices." Instead, these patterns likely reflect adaptive strategies to the complexities of different team environments, offering distinct takeaways for various team structures.

\textbf{High Collaboration Environments:}
Projects with high collaboration intensity exhibit a fascinating and complex dynamic. Their overall commit activity is correlated with a higher volume of artifact-producing commits like \textit{Output Data}, \textit{Internal Documentation}, and \textit{Sharing} (Finding 4). However, their commit-to-commit evolution shows a significant increase in transitions to \textit{External Documentation} and a decrease in transitions to \textit{Output Data} (Finding 11). Meanwhile, their formal releases are more likely to involve sequences of technical work like \textit{Training Infrastructure}, \textit{Model Structure}, and \textit{Parameter Tuning} (Finding 14).
This suggests a sophisticated workflow in collaborative projects:
\begin{enumerate}
    \item The day-to-day churn involves generating many artifacts and internal logs for tracking.
    \item However, the communicated evolution (commit sequences) prioritizes shared understanding via external documentation over simply checking in new individual results.
    \item Major technical advancements are then bundled and consolidated into formal, technically-dense releases.
\end{enumerate}
This separation of concerns, using commits for communication and releases for technical milestones, is a key adaptive strategy for managing complexity in collaborative ML development.

\textbf{Implications for Different Team Settings:}
\begin{itemize}
    \item \textbf{Individual Contributors/Small Teams:} The prominence of \textit{Output Data} and \textit{Project Metadata} management (Finding 1) implies that even solo developers must adopt MLOps-like discipline early. They can learn from the patterns in collaborative projects: while generating artifacts is core, consciously creating \textit{External Documentation} (even if for their future selves) and using releases to bundle technical milestones (Finding 14) rather than just new weights can improve project sustainability and reproducibility.
    \item \textbf{Large Groups/Organizations:} For these teams, our findings reinforce the critical role of managing the communication-artifact trade-off revealed in Finding 11. They must establish clear protocols for documentation to manage dependencies, as their natural workflow prioritizes this. The finding that their releases are technically dense (Finding 14) should guide their release planning to be centered around significant technical advancements, not just routine model updates. The patterns in popular projects (Finding 7), which often have organizational backing, suggest that establishing project templates to streamline metadata setup can allow teams to more quickly focus on scalable infrastructure and dissemination.
\end{itemize}
This tailored approach, considering team structure and project goals, can help translate observed patterns into more actionable development strategies for ML projects.

\section{THREATS TO VALIDITY MITIGATIONS}
\label{sec:threats}

In designing our methodology, we proactively addressed potential threats to the validity of our study to ensure the robustness and reliability of our findings. The following mitigations were implemented:

\textbf{Construct Validity:} To ensure that our measures accurately capture the constructs of interest, we employed rigorous data cleaning and preprocessing procedures. The use of the Gemini 2.5 Flash LLM for commit classification was validated through manual checks, achieving high accuracy and ensuring that the classification comprehensively represents the complexity of commit types and popularity metrics.

\textbf{Conclusion Validity:} We utilized established statistical methods, including BNs and DBNs, to model the complexities of model evolution. Thorough validations of our models and sensitivity analyses were conducted to ensure the robustness of our findings. These methodologies were informed by existing literature, thereby enhancing the credibility of our conclusions.

\textbf{Internal Validity:} To mitigate biases in commit classification and analysis, we employed a proven methodology from prior research. Validation checks, including manual analysis of a subset of commit messages, ensured high alignment with automated classifications. This approach minimized the influence of biases inherent in the training data or model architecture.

\textbf{External Validity:} Recognizing the limitation of focusing solely on the HF platform, we ensured that our methodology is robust and replicable, allowing application to future datasets or similar platforms. Providing a detailed methodology and a replication package facilitates validation of our findings across different contexts, thereby enhancing the generalizability of our results. Furthermore, our analysis of releases (RQ3) specifically targets models using explicit Git tags for versioning. As our data collection indicated that only a subset of models on HF employed detectable tags (1,655 models found initially before filtering for release analysis), the findings related to release patterns might not generalize to models using implicit or alternative versioning strategies. The RQ3 results are most representative of projects adopting standard Git-based release management practices on the platform. Finally, it is important to acknowledge the known domain imbalance on HF, which heavily favors Natural Language Processing (NLP) models, as also highlighted by \citet{jiang2024peatmoss}. Consequently, while our study includes analyses across different identified domains (RQ1.2), our aggregate findings and broadly trained Bayesian Network models may disproportionately reflect practices within the NLP community. Caution should therefore be exercised when generalizing specific quantitative results or evolutionary patterns too broadly to less represented domains, such as Reinforcement Learning or particular subfields of Computer Vision and Audio, without further dedicated, domain-specific investigations.\par A further limitation is the absence of a demographic analysis of the model authors in our sample. Understanding the distribution of models contributed by organizations versus individual users would provide valuable context for our findings on development patterns and community practices. We acknowledge this as an important dimension for future research.

\textbf{Reliability:} To address potential changes in the HF API or the underlying Git repository structures it interfaces with, we comprehensively documented all steps of our data collection (including direct Git processing for commit file lists) and preprocessing methods. The replication package includes detailed instructions and the necessary code to reproduce our study, ensuring that future researchers can replicate our findings despite potential underlying data source changes.

\section{CONCLUSIONS AND FUTURE WORK} \label{sec:conclusions}

In this study, we conducted a large-scale analysis of how ML models evolve over time within the open-source ecosystem, focusing on the HF platform. Some of the key contributions are:

\begin{enumerate}
    \item \textbf{Large-Scale Classification of Model Changes:} We applied and extended an ML change taxonomy to classify over 680,000 commits across   100,000 models on HF, identifying prevalent commit types ( \textit{Output Data}, \textit{Project Metadata}, and \textit{Sharing}) and their distribution throughout project lifecycles.
    \item \textbf{Uncovering Patterns in Commit and Release Activities:} Utilizing BNs and DBNs, we identified powerful temporal patterns in commit and release sequences, revealing the underlying workflows and dependencies that drive model development.
    \item \textbf{Insights into Model Evolution, Popularity, and Collaboration:}  Our analysis showed that popular projects often start from a more mature baseline and evolve by prioritizing performance and dissemination over raw artifact generation (Findings 7, 12). Furthermore, projects with high collaboration intensity exhibit a sophisticated workflow, using commit sequences for communication and formal releases for consolidating technical advancements (Findings 11, 14).
    \item \textbf{Distinction Between Commits and Releases:} We found that releases tend to consolidate significant updates in  \textit{Output Data}, \textit{Sharing}, and \textit{External Documentation} (Finding 13), serving as key communication milestones. This distinguishes their role from the more granular, frequent changes seen in commits, which are dominated by \textit{Output Data}, \textit{Project Metadata}, and \textit{Sharing} (Finding 1).
    \item \textbf{Alignment with CRISP-DM and Clustering of Activities:} We demonstrated that model changes on HF reflect iterative development processes alignable with frameworks like CRISP-DM. This is supported by the phased nature of commit activities (Finding 3) and the strong bundling of disparate tasks into single, cohesive experimental commits (Finding 8).
\end{enumerate}

These contributions enhance our understanding of model maintenance and improvement practices on community platforms, offering valuable guidance for best practices in model development and management.

Our findings have concrete implications for the software engineering community, particularly when developing and maintaining ML models, which present unique challenges and evolutionary patterns distinct from much traditional software:

\begin{itemize}
    \item \textbf{Tailoring Maintenance and Operations for ML Systems:} The observed prevalence of \textit{Output Data} and \textit{Project Metadata} commits (Finding 1) highlights that ML model maintenance is not solely about code. It critically involves managing evolving data products and complex repository configurations. This calls for robust MLOps strategies that treat data artifacts, model weights, and project configurations as first-class versioned entities, on par with source code.
    
    \item \textbf{Optimizing ML's Inherently Experimental Workflow:} The iterative and empirical nature of ML development is underscored by the powerful "work -> artifact" cycle seen in commit sequences (Finding 9) and the extreme bundling of tasks into single commits (Finding 8). This suggests that development proceeds in atomic experimental units. Recognizing this, teams can structure projects and adopt tooling that manages and tracks these multi-faceted experiments holistically, rather than as a series of disconnected code, parameter, and data changes.
    
    \item \textbf{Adapting Collaboration and Documentation for the ML Lifecycle:} The distinct patterns linked to collaboration intensity reveal a sophisticated, multi-layered communication strategy. The focus on documentation transitions in day-to-day commits (Finding 11) alongside technically-dense releases (Finding 14) implies that collaborative ML projects separate continuous alignment (via commit-level communication) from major technical consolidation (via releases). This highlights the need for clear documentation protocols to manage dependencies and a deliberate release strategy centered on communicating significant technical advancements.
\end{itemize}

\paragraph{Future Work}

Future research could extend our analysis by exploring the impact of identified commit and release patterns on model performance and user adoption, thereby linking development practices with tangible outcomes. Additionally, applying our taxonomy and analytical framework to other platforms such as GitHub or domain-specific repositories would help validate the generalizability of our findings across the broader ML ecosystem. Developing automated tools that leverage our classification and pattern recognition techniques could assist developers in adopting best practices for model maintenance and collaboration. Moreover, longitudinal studies tracking models over longer periods could provide deeper insights into the sustainability and evolution of successful models, while investigating the role of security and compliance in model updates would address critical aspects of operational sustainability. Integrating our findings with MLOps practices could further enhance continuous integration and deployment workflows, ensuring models remain up-to-date and performant in dynamic production environments.

\section{ACKNOWLEDGMENT}
This work was primarily supported by GAISSA (TED2021-130923B-I00), funded by MCIN/AEI/10.13039/5011000110 33 and the European Union NextGenerationEU/PRTR. Additionally, J.C.\ acknowledges funding from the FI-STEP predoctoral fellowship (grant no.\ 2025 STEP 00414) awarded by AGAUR (Government of Catalonia) and co-funded by the European Social Fund Plus (FSE+) of Catalonia 2021--2027; A.S.\ and R.C.\ acknowledge support from PID2022-139293NB-C31, funded by MCIN/AEI/10.13039/501100011033 and by ERDF (``A way of making Europe''). R.C.\ also acknowledges support from the Spanish Ministry of Science, Innovation and Universities through the ``Mar\'{i}a Zambrano'' grant (RR\_C\_2021\_01) funded by NextGenerationEU, and from the ``Plan Propio de Investigaci\'{o}n y Transferencia 2024--2025'' of the University of Almer\'{i}a (project P\_LANZ\_2024003).

\bibliographystyle{ACM-Reference-Format}
\bibliography{References}

\end{document}